  \documentclass[journal,onecolumn,draftcls,12pt]{IEEEtran}
\usepackage{cite}
\usepackage{multirow}
\usepackage{graphicx}
\usepackage{amssymb,amsmath,times,wasysym,amsthm}
\usepackage{pb-diagram}
\usepackage{psfrag,balance}
\usepackage{srcltx,verbatim} 

\usepackage{fixltx2e}
\usepackage{multicol}
\usepackage{xcolor}
\usepackage{algorithmic}
\usepackage{algorithm}
\usepackage[caption=false]{subfig}

\newcounter{appendx}

\newcommand{\real}[1]{\Re\left({#1}\right)}

\renewcommand{\exp}[1]{\mathrm{e}^{#1}}

\renewcommand{\log}[2][]{\mathrm{log}_{#1}\left(#2\right)}

\newcommand{\Q}[1]{\mathcal{Q}\left(#1\right)}

\newcommand{\GammaFn}[1]{\Gamma\!\left({#1}\right)}

\newcommand{\E}[2][]{\mathbb{E}_{#1}\!\left[{#2}\right]}

\newcommand{\e}[1]{\text{exp}\!\left({#1}\right)}

\newcommand{\Var}[1]{\mathbb{V}\mathrm{ar}\!\left[{#1}\right ]}

\title{{Performance Analysis of Intelligent Reflective Surface Aided Wireless Communications} \vspace{-0mm}}

\author{ Dulaj Gunasinghe, \IEEEmembership{Student Member, IEEE}, Dhanushka Kudathanthirige, \IEEEmembership{Student Member, IEEE}, and  {Gayan Amarasuriya Aruma Baduge, \IEEEmembership{Senior Member, IEEE}}
	
	\thanks{The authors are with the Department of Electrical and Computer Engineering, Southern Illinois University, Carbondale, IL, USA, Email: \{dulaj.gunasinghe,dhanushka.kudathanthirige,gayan.baduge\}@siu.edu. This work in part has been presented at IEEE International Conference on Communications (ICC), June, 2020 \cite{Kudathanthirige2020}.} \vspace{-10mm}}

\DeclareMathAlphabet\mathbfcal{OMS}{cmsy}{b}{n}

\begin{document}
\bstctlcite{IEEEexample:BSTcontrol}
\vspace{-0mm}
\maketitle
 
\begin{abstract}
	\vspace{-0mm}

	The fundamental performance metrics of an intelligent reflective surface (IRS)-aided wireless system are presented. By optimizing the IRS phase-shift matrix, the received signal-to-noise ratio (SNR) is maximized at the destination in the presence of both reflected and direct channels. The probability distributions of this maximum SNR are tightly approximated for the moderate-to-large reflective element regime. Thereby, the probability density function  and cumulative distribution function  of this tight SNR approximation are derived in closed-form for Nakagami-$m$ fading to facilitate a statistical characterization of the performance metrics. The outage probability, average symbol error probability, and achievable rate bounds are derived. By virtue of an asymptotic analysis in the high SNR regime, the diversity order is quantified.  Thereby, we reveal that the overall diversity order can be scaled as a function of the number of reflective elements ($N$) such that $G_d = m_v + \min(m_g,m_h)N$, where $m_v$, $m_h$ and $m_g$ are the Nakagami-$m$ parameters of the direct, source-to-IRS and IRS-to-destination channels, respectively. The asymptotic achievable rate is derived, and thereby, it is shown that the transmit power can be scaled inversely proportional to $N^2$. The impact of quantized IRS phase-shifts is investigated by deriving the achievable rate bounds. Useful insights are obtained by analyzing the system performance for different severity of fading cases including spatially correlated fading. Our   analysis and numerical results reveal that IRS is a promising technology for boosting the performance of   wireless communications by intelligently controlling the propagation channels without employing additional active radio-frequency chains.

\end{abstract}

\newpage 
 
 \linespread{0}

 
\section{Introduction}\label{sec:introduction}
\vspace{-0mm}

Over the past five generations of wireless standards,  performance of the  transmitter and receiver has been optimized 
to mitigate various transmission impairments of propagation channels, which are generally assumed to be uncontrollable in  the wireless system designer's  perspective. However, owing to the recent research advancements of meta-materials and meta-surfaces,  a novel concept of coating physical objects such as building walls and windows with intelligent reflective surfaces (IRSs) with reconfigurable  reflective  properties has  been envisioned \cite{Kudathanthirige2020,Liaskos2018,Renzo2019}. The ultimate goal of IRS is to enable a smart wireless propagation  environment by controlling the reflective properties of the  underlying channels \cite{Renzo2019}.

   An IRS comprises of  a very large number of passive reflective elements, which are capable of  reconfiguring  properties of electromagnetic (EM) waves impinging upon them.    On one hand, reflected EM waves can be added constructively at a  desired  receiver by intelligently controlling  phase-shifts at each reflective element to boost the signal-to-noise ratio (SNR) and coverage. On the other hand, a   reflected  signal can be made to    add  destructively and thereby to mitigate co-channel interference towards an undesired direction.
  Moreover, IRS   facilitates full-duplex reflections, and hence, large blockages between a pair of transmitter-receiver can be circumvented through smart reflections without trading-off additional time, frequency or power resources.       
  Since an IRS does not generate new EM waves, costly transmit radio-frequency (RF) chains/amplifiers in relays can be eliminated and thereby improving the energy efficiency. 
  Thus, the concept of IRS presents  a paradigm shift in wireless communication research.

  \vspace{-2mm}
  
   \subsection{A literature survey on intelligent reflective surfaces for wireless applications} 
       	\vspace{-2mm}
       	
    The fabrication  of software-controllable  IRS has been shown to be feasible owing to  the recent breakthroughs  in physics and related fields \cite{Lee2012}.  The     core technical aspects of modeling IRS to enable reconfigurable EM properties  are currently being developed \cite{Liaskos2018}. 
    The prototypes of meta-surfaces and meta-tiles with artificial thin-film of EM  materials, which are intended to  coat physical objects to enable a   smart wireless environment, are being  developed \cite{Lee2012}. 
    
    Recently,    several attempts of adopting IRS into wireless system designs have been reported \cite{Wu2019, Basar2019, Han2019, Chen2019, Zhang2020,Abeywickrama2020, Jung2020,Nadeem2019,Zhang2019,Jung2019b,Badiu2020,Psomas2019,Guo2020,Basar2019b,Hu2020,Zhang2020b}.
        The joint precoder      and IRS   phase-shift  optimization techniques to maximize the received SNR are investigated in \cite{Wu2019} for a multi-antenna transmitter in the presence of an IRS.  Reference   \cite{Basar2019} adopts  basic ray tracing techniques  to model multi-path propagation through an IRS, and thereby, authors propose transmission strategies to control the reflections by virtue of   phase-shift optimization at passive elements embedded within an IRS. 
        In \cite{Abeywickrama2020}, joint IRS and beamforming optimization is studied with a practical phase-shift model.    
             Reference  \cite{Han2019} proposes    an efficient phase-shift optimization  design  
   based on maximizing an  upper bound of the average spectral efficiency of the IRS-aided communications. 
The smart propagations enabled by an IRS have been exploited in \cite{Chen2019} to boost the  physical layer security.
   	   	In \cite{Zhang2020},  the impact of having a limited number of phase-shifts at the IRS and the corresponding phase-shift designs are investigated.  In \cite{Jung2020}, the asymptotic achievable  uplink   rate  is derived by adopting the  channel hardening effects for very large  {reflective} element regime.    Reference \cite{Nadeem2019} investigates optimization techniques to maximize the signal-to-interference-plus-noise ratio (SINR) of an IRS-aided communication between a  multi antenna base-station (BS) and  single-antenna users. 
   	In \cite{Zhang2019},  the outage probability is studied  for multiple IRS-aided wireless systems. 
   	In \cite{Jung2019b}, the asymptotic rate distribution of the IRS-aided systems operating in Rician fading is derived.  
   	In \cite{Badiu2020},  the IRS-based communication with phase-shift errors is investigated, and  it has been shown that the cascaded channel through an IRS with phase errors is equivalent to a point-to-point channel with Nakagami fading.
   	The performance of a  random rotation-based IRS communication is studied  in \cite{Psomas2019} by proposing four low-complexity and energy efficient techniques via   the coding- and selection-based approaches. 
   	In \cite{Guo2020}, the maximal ratio transmission   precoder is used at the BS to facilitate IRS-based communication with  single-antenna users, and thereby, the outage probability minimization  is performed  by optimizing the IRS phase-shifts.
   	   	In \cite{Basar2019b}, the authors have shown that the multi-path fading   and the related Doppler effects 	caused by the relative movement of the mobile nodes can be    mitigated by real-time adjustable IRS phase-shifts. In \cite{Basar2019b}, a number of design trade-offs   between the fade pattern elimination and complex	envelope magnitude maximization has been investigated. 
   	   	In \cite{Diluka2020}, the impact of distributed IRS deployments is studied.    	   	
   	   	 In \cite{Hu2020}. 
   	it has been shown that when compared with the two-dimensional (2D) IRS, the spherical IRS has several advantages such as wider coverage, simpler positioning technique and flexible deployment. 
   	In \cite{Zhang2020b},  the capacity region of an IRS-aided two user multiple access channel has been  characterized for distributed and centralized IRS deployment, and thereby,  it has been shown that centralized deployment outperforms the distributed counterpart.

   	This paper goes beyond our related conference paper \cite{Kudathanthirige2020} by presenting the performance metrics with the direct channel,  Nakagami-$m$ fading, asymptotic   analysis in large reflective element regime, transmit power scaling laws, and impact of phase-shift quantization. In  \cite{Kudathanthirige2020}, Rayleigh fading has been considered, and the outage probability, average {symbol error rate (SER)} and achievable rate of  the IRS reflected channel  (without the direct channel) have been derived.

   	\vspace{-3mm}
   
    \subsection{Motivation and our contribution} 
       	\vspace{-3mm}

    The main contribution of this paper is to present a closed-form performance analysis for IRS-aided wireless systems operating  over Nakagami-$m$ fading. 
      The key idea of an IRS is  to enable a programmable control over the wireless propagation channels.      This necessitates  innovations of  radically  novel techniques for modeling, designing and analyzing wireless systems as the resulting  smart propagation channels can now be able to interact with    EM waves impinging upon them in a software-controlled manner. Although  several important attempts  have recently been made 
      \cite{Liaskos2018,Renzo2019,Lee2012,Wu2019, Basar2019, Han2019, Chen2019, Zhang2020, Jung2020,Nadeem2019,Zhang2019,Jung2019b,Badiu2020,Psomas2019,Guo2020,Basar2019b,Hu2020,Zhang2020b}, the fundamental research on IRS in wireless communication's perspective is still at an embryonic stage. For instance, a mathematically tractable statistical characterization of the optimal end-to-end SNR that is attainable via an IRS-aided reflections and direct channel over Nakagami-$m$ fading has not yet been investigated. Specifically, Nakagami-$m$ fading model is versatile in the sense that it can model a wide-range of multi-path fading environments \cite{Stuber2017}. For instance,  $m=1/2$ captures the most severe fading with {a} one-sided Gaussian distribution, while $m=1$ models the Rayleigh fading with rich-scattering. When $m\approx(K+1)^2/(2K+1)$, Nakagami-$m$ fading can be used to approximate Rician fading, where $K$ is the Rician factor \cite{Stuber2017}. It can  also capture less severe fading cases when $m$ is large.
      Moreover, insightful  closed-form analyses on the outage probability, average {SER}, achievable rate, diversity order and array gain have not yet been reported to IRS-aided communications in the presence of the direct channel. The detrimental impact of phase-shift quantizations at the IRS elements  has  not yet been analytically quantified into the above performance metrics. The above facts and the important research gaps have motivated the current work.      
       To this end, our paper presents a comprehensive  performance analysis framework for deriving the fundamental bounds pertaining to an IRS intended for aiding the end-to-end communication between a single-antenna source ($S$) and a destination ($D$). The distinct  contribution of this work can be summarized as follows:
      \begin{enumerate}
      	\item We statistically characterize the  received SNR that is attainable by virtue of optimal phase-shift control at the IRS elements such that all reflected channels are constructively combined with the direct channel. To this end,  the probability density function (PDF) and cumulative function (CDF) of {a} tight approximation of the optimal SNR are derived in closed-form for Nakagami-$m$ fading. We reveal that this statistical characterization is mathematically tractable to facilitate   the closed-form derivations of important performance metrics, and the resulting PDF/CDF approximations are significantly tight to the exact counterparts in the moderate-to-large {reflective} element regime.     
      	

      	
      	\item By using our PDF/CDF analysis, novel, tight approximations/bounds     for the outage probability, average SER, and achievable rate are derived in closed-form. Thereby,  useful design insights are obtained. By using the  asymptotic  outage probability and average SER analyses in the high SNR regime,  the  diversity order and array/coding gain are quantified. This  asymptotic analysis reveals that the  diversity order is an affine function of the number of IRS elements ($N$), and hence, the system reliability metrics  such as the outage probability and average SER can be  boosted merely by virtue of passive reflections of exiting EM waves without adopting to costly/active RF chains.  Moreover, we derive the asymptotic achievable rate in the large {reflective} element regime ($N \rightarrow \infty$), and thereby, we reveal that the our lower and upper rate bounds are asymptotically exact as $N\rightarrow \infty$. Specifically, we show that the transmit power can be scaled inversely proportional to the square of the number of IRS elements in the asymptotic $N$ regime, while   achieving a finite rate. 
      	
 \item Finally, we discuss the practical design aspects and draw   insights via our analysis. In this context, we investigate the impact of hardware-limited quantized phase-shifts at the IRS elements. The underlying deleterious effects are analytically quantified by deriving the outage probability and achievable rate bounds. Thereby, we reveal that the {quantized} IRS phase-shifts considerably hinder the system performance compared to the continuous phase-shifts. Further, we show that our analysis can be used to deduce performance metrics for distinctly asymmetric fading cases including, the most severe one-sided Gaussian, Rayleigh/rich-scattering,   unserviceable direct channel,  and less severe multi-path fading. Moreover, the impact of spatially correlated fading is studied.  
      	
  
      \end{enumerate}


   \begin{figure}[!t]\centering \vspace{-5mm}
   	\def\svgwidth{190pt} 
   	\fontsize{8}{4}\selectfont 
   	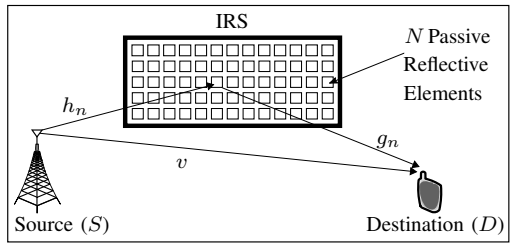 \vspace{-1mm}
   	\caption{An IRS-aided  wireless   set-up. The distances  $S$-$D$, $S$-IRS, and  $D$-IRS are denoted by $d_{SD}$, $d_{SI}$, and $d_{DI}$, respectively.}\vspace{-10mm} \label{fig:system_model}
   \end{figure}

 \noindent \textbf{Notation:} 
  $\E[]{X}$ and $\Var{X}$ are the expectation and variance of  a random variable (RV) $X$, respectively.
  $X\sim\mathcal {N}\left(\mu_X,  \sigma_X^{2}  \right) $ denotes that $X$ is  Gaussian distributed with $\mu_X$ mean and $\sigma_X^{2}$ variance.  $\GammaFn{z}=\int_{0}^{\infty} \exp{-t}t^{z-1}dt$ is the Gamma function \cite[8.310.1]{Gradshteyn2007}.      $\Gamma(q,z)=\int_z^\infty t^{q-1}\exp{-t}dt$  \cite[8.350.2]{Gradshteyn2007} and $\gamma(q,z)=\int_0^z t^{q-1}\exp{-t}dt$ \cite[8.350.1]{Gradshteyn2007} are the upper and lower incomplete Gamma functions, respectively.
   $\mathcal O(z^p)$ denotes the remainder in a Maclaurin series \cite[0.318.2]{Gradshteyn2007}   after the $z^p$ term, and {$\otimes$ denotes the Kronecker product.}

 \vspace{-2mm}
 \section{System, channel and signal models}\label{sec:system_model}

 
 \vspace{0mm}
\subsection{System and channel models}\label{sec:channel_model}
\vspace{-2mm}

We consider an IRS-assisted  communication system in which a source $(S)$ communicates with a destination $(D)$ via    an IRS   (Fig. \ref{fig:system_model}). The $S$ and $D$ are single-antenna nodes, while the IRS is embedded with  $N$-passive reflective elements. The  IRS is  able to control the phase-shifts of  incident/impinging EM waves such that the received signals can be constructively  combined at $D$, while the reflected signal  amplitudes are also attenuated. Thus, the reflective coefficient of the  $n$th IRS element can be modeled as {$r_n =\eta_n \exp{j\theta_n}$}, where $0<\eta_n\leq 1$ and $-\pi<\theta_n\leq \pi$ are amplitude attenuation coefficient and phase-shift, respectively, for $1\leq n\leq N$.   

The channels from $S$ to $D$,   $S$ to the $n$th element of IRS, and   the $n$th IRS element to $D$ are denoted by $v$, $h_n$  and  $g_n$, respectively, where $1\leq n\leq N$. These complex  channel coefficients can be written in polar form as $v = \bar{v} \,\e{j \phi_{v}}$, $h_n= \bar{h}_n \e{j \phi_{{h}_n}},\text{ and } g_n = \bar{g}_n \e{j \phi_{{g}_n}}$, where $\{\bar{v}, \bar{h}_n,\bar{g}_n\}$ and $\{\phi_{v}, \phi_{{h}_n}, \phi_{{g}_n}\}$ are the channel amplitudes/envelopes and phases of the respective channels. To represent a wide range of fading scenarios,  we adopt Nakagami-$m$ channel fading model. Therefore, the PDF of
$\bar{v}$ is given by 
\begin{eqnarray}\label{eqn:Nakagami_v}
f_{\bar{v}}(x) = \frac{2m_v^{m_v} x^{2m_v-1}}{\Gamma(m_v) \kappa_{v}^{m_v}} \e{\frac{- m_vx^2}{\kappa_{v}}} \text{ for } x\geq 0,
\end{eqnarray}
where $m_v$ and $\kappa_{v}=m_v\zeta_{v}$, respectively, denote the shape and scaling parameters of the Nakagami-$m$ distribution \cite{Papoulis2002}. Here, $\zeta_v$ captures  the large-scale fading of the channel. 
Similarly, the PDFs of  $\bar{g}_n$ and $\bar{h}_n$ can be readily obtained by replacing  $m_v$ and $\kappa_{v}=m_v\zeta_{v}$ in (\ref{eqn:Nakagami_v}) by the  corresponding channel parameters;  $\{m_g$ and $\kappa_{g}=m_g\zeta_{g}\}$ and $\{m_h$ and $\kappa_{h}=m_h\zeta_{h}\}$.

 \vspace{-0mm}
 \subsection{Signal model}
 
 \vspace{-2mm}
The   signal received at $D$ via the  direct   and IRS-aided reflected channels can be written as \cite{Wu2019}
\vspace{-8mm}
	 \begin{eqnarray}\label{eqn:received_signal}
	r &=&\sqrt{p} v x+\sqrt{p} \displaystyle \sum_{n=1}^{N} g_n \eta_n \e{j\theta_n} h_n x+ n,
	\end{eqnarray}  
	
	\vspace{-2mm}
	\noindent
where $p$ is  transmit power, and $x$ is transmitted signal at $S$, satisfying $\E{|x|^2}=1$.  Moreover, $n$ is an additive white Gaussian noise (AWGN) at $D$ with zero mean and variance $\sigma^2$  such that $n\sim \mathcal {CN}(0, \sigma^2)$. In (\ref{eqn:received_signal}), the first and second components account for the received signal via the direct  channel and IRS-aided reflected channel, respectively. 
By replacing $v$, $h_n$, and $g_n$ by their  polar forms presented in  Section \ref{sec:channel_model}, the received SNR at $D$ can be derived by using (\ref{eqn:received_signal}) as

\vspace{-8mm}
 \begin{eqnarray}\label{eqn:SNR}
\tilde{\gamma} = { \bar{\gamma}\Big| \bar{v}\e{j\phi_v}+\displaystyle \sum_{n=1}^{N} \bar{g}_n \bar{h}_n \eta_n \e{[j (\phi_{g_n}+\phi_{h_n}+\theta_n)]} \Big| ^2 },
\end{eqnarray}

\vspace{-2mm}\noindent
 where $\bar{\gamma}=p/\sigma^2$ is defined as the transmit SNR.
 
 \vspace{10mm}
 \noindent \textbf{Lemma 1:}\textit{The maximum received SNR at $D$ that can be attained by controlling the phase-shifts at the IRS elements is given by }
 
 \vspace{-10mm}
  \begin{eqnarray}\label{eqn:SNR_Optimal}
 \tilde\gamma^* =   { \bar{\gamma}\left| \bar{v} + \displaystyle \sum_{n=1}^{N} \bar{g}_n \bar{h}_n \eta_n \right| ^2 },
 \end{eqnarray}
 \textit{and the corresponding optimal phase-shifts are given by} 
 \vspace{-4mm}
 \begin{eqnarray}\label{eqn:optimal_theta}
  \theta^*_n = \phi_v-(\phi_{h_n}+\phi_{g_n}), \;\;\; \text{for } 1\leq n\leq N.
 \end{eqnarray}
 \vspace{-12mm}
 
	 \linespread{1.6}
\begin{proof}
 The signals received at $D$ via the IRS-aided reflected channel or equivalently $N$ terms inside the summation of (\ref{eqn:SNR}) must be constructively added to the signal received via the direct channel   to maximize the received SNR at $D$. This can be accomplished by controlling the phase-shift of each IRS element ($\theta_n$) to match the phase of the direct channel ($\phi_v$), i.e., $\phi_v=\phi_{g_n}+\phi_{h_n}+\theta_n,\; \forall n$. Thereby, the optimal phase-shift at the $n$th IRS element can be derived as shown in (\ref{eqn:optimal_theta}). By substituting (\ref{eqn:optimal_theta}) into (\ref{eqn:SNR}), the maximum SNR at $D$ can be derived as (\ref{eqn:SNR_Optimal}).
\end{proof} 
 
 	 \linespread{0}
%
%

\vspace{-3mm}

\section{Performance Analysis}\label{sec:performance_analysis}

\begin{figure*}[!t]\centering \vspace{-5mm}
	\subfloat[ PDF]{\hspace{-0mm}
		\includegraphics[width=0.4\textwidth]{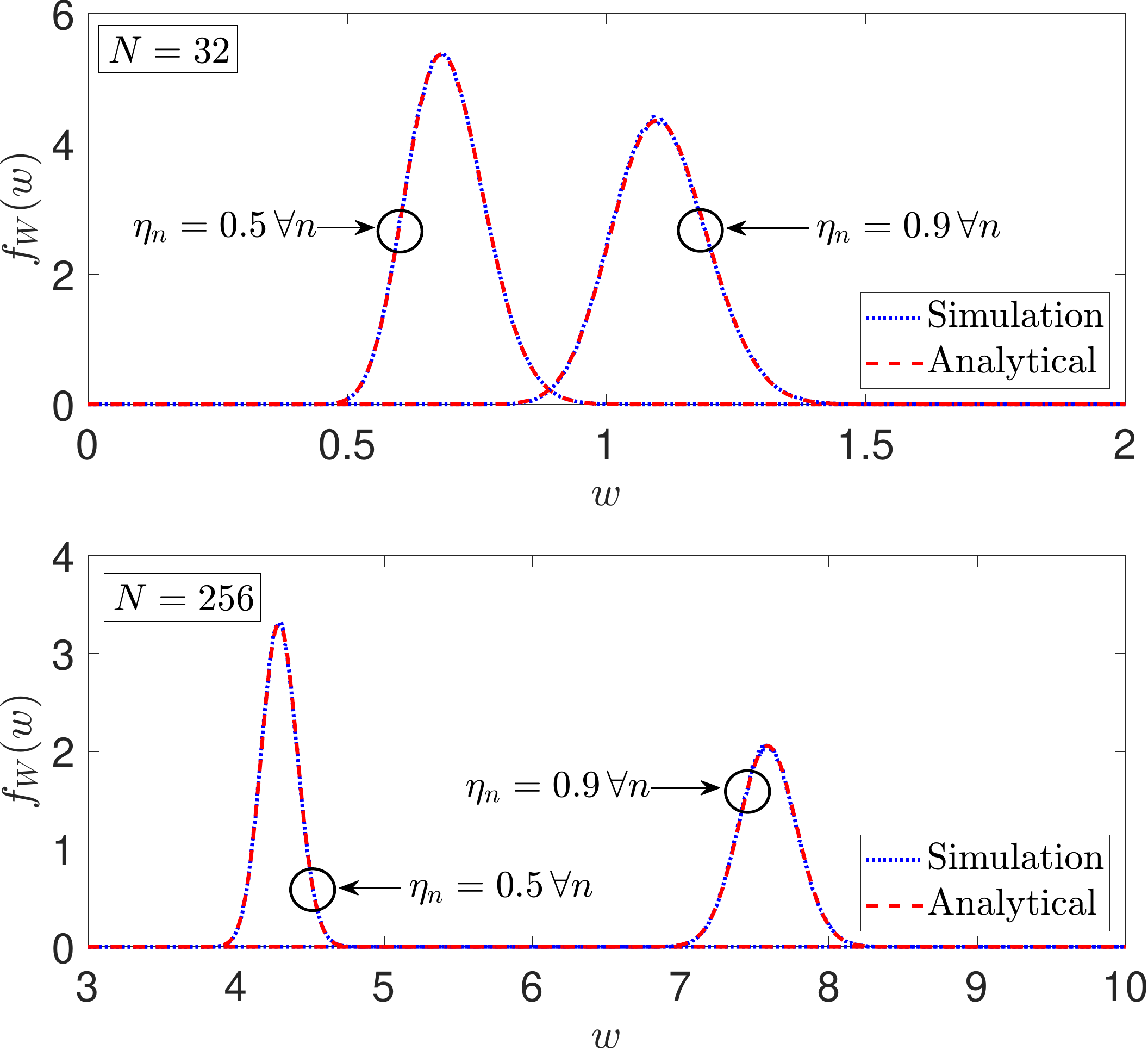}   \label{fig:PDF} }
	\subfloat[CDF]{\hspace{-0mm}
		\includegraphics[width=0.4\textwidth]{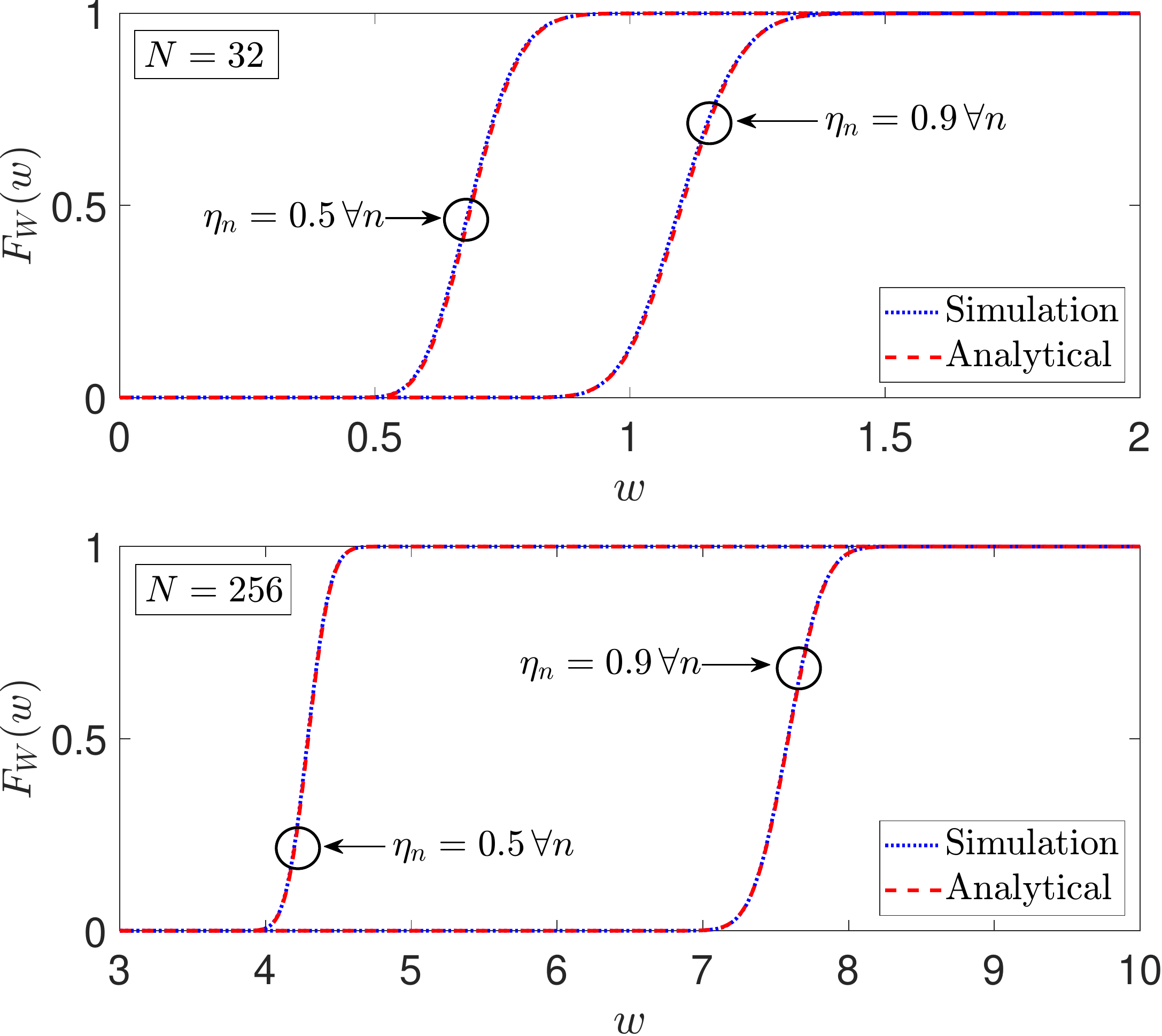}
			\label{fig:CDF}  }\vspace{-2mm}
	\caption{The exact PDF and CDF of $\tilde W=\sum_{n=1}^{N} \bar{g}_n \bar{h}_n \eta_n$ via Monte-Carlo simulations and  our analysis (\ref{eqn:pdf_S}) and (\ref{eqn:CDF_S}).}\vspace{-10mm}
	\label{fig:Fig23} 
\end{figure*}

%

\subsection{Statistical characterization of the optimal received SNR}

To begin with, we derive a tight approximation for $ \tilde\gamma^*$ in (\ref{eqn:SNR_Optimal}). To begin with, we notice that
\begin{eqnarray}\label{eqn:W_tilde}
 \tilde{W}=\sum_{n=1}^{N} \bar{g}_n \bar{h}_n \eta_n
\end{eqnarray}
  is a sum of the products of independent Nakagami distributed RVs  $\bar{g}_n$ and $\bar{h}_n$ for $n\in \{1,\cdots, N\}$. The derivation of the exact probability distribution of $\tilde{W}$ is mathematically involved and may not provide   useful technical insights.
Thus, we resort to   tight approximations to facilitate a mathematically tractable performance analysis.  To this end, by invoking  the central limit theorem (CLT) \cite{Papoulis2002},  $\tilde W$ in (\ref{eqn:W_tilde}) can be tightly approximated  by  $ W$, whose PDF and    CDF are presented in Lemma 2.

\noindent \textbf{Lemma 2}: \textit{For a sufficiently large number of   reflective elements, the PDF and CDF of $W$ are given by} 
\vspace{-5mm}
\begin{eqnarray}
f_W(w)=\frac{\xi}{\sqrt{2\pi}\bar{\sigma}}\e{-(w-\bar{\mu})^2/2\bar{\sigma}^2}\;\; \text{ for } w\geq 0 \textit{ and } 0 \textit{ otherwise}, \label{eqn:pdf_S}\\
F_W(w) =1- \xi\Q{\left({w-\bar{\mu}}\right)\Big/ {\bar{\sigma}}} \text{ for } \quad w \geq 0 \textit{ and } 0 \textit{ otherwise}, \label{eqn:CDF_S} 
\end{eqnarray}
 \textit{where  the parameters   $\bar{\mu}$ and $\bar{\sigma}^2$   can be defined  as }
 	\begin{eqnarray}
 	\bar{\mu}&=&\sum_{n=1}^N\eta_n\sqrt{\frac{\kappa_{g_n}\kappa_{h_n}}{m_gm_h}}T(m_g,m_h,1/2),\label{eqn:meanbar}
 	\text{ and }
 	\bar{\sigma}^2=\sum_{n=1}^N\eta^2_n\kappa_{g_n}\kappa_{h_n}\left(1-\frac{T^2(m_g,m_h,1/2)}{m_gm_h}\right),
 	\label{eqn:variancebar}
 	\end{eqnarray}
 	\textit{where $T(a,b,i)={\Gamma(a+i)\Gamma(b+i)}\big/\left({\Gamma(a)\Gamma(b)}\right)$, and  $\mathcal{Q}(\cdot)$ denotes the Gaussian-$\mathcal{Q}$ function with  $\xi=1/\mathcal Q (\bar{Z})$ and $\bar{Z}=-\bar{\mu}/\bar{\sigma}$.}
\begin{proof}
	See Appendix  \ref{app:AppendixA1}.
\end{proof}
%
%

\textit{\textbf{Remark 1:}}
The accuracy of {Lemma 2} is   verified via Monte-Carlo simulations 
in Fig. \ref{fig:PDF} and Fig. \ref{fig:CDF}, where our  analytical      PDF and CDF  of $W$ in (\ref{eqn:pdf_S}) and (\ref{eqn:CDF_S}) are plotted   for different $N$ values. These figures reveal that our analytical PDF and CDF are significantly tight to  the exact Monte-Carlo simulations even for moderately large $N$ values. 

\textit{\textbf{Remark 2:}}
By using Lemma 2, it can be shown that as $N$ grows without bound, the mean and variance of $W$ indeed approach   $\bar{\mu}$  and $\bar{\sigma}^2$ in (\ref{eqn:variancebar}). 
To prove this, we first derive the mean and variance of $W$ by using the PDF in (\ref{eqn:pdf_S}) as follows \cite{Papoulis2002}:
\vspace{-3mm}
\begin{eqnarray}
\mu_{W}&=&\bar{\mu}+\bar{\sigma}\xi\phi(\bar{Z}) \label{eqn:meanS} \quad \text{and}\quad 
\sigma^2_{W}=\bar{\sigma}^2\left(1+\bar{Z}\xi\phi(\bar{Z})-(\xi\phi(\bar{Z}))^2\right),
\label{eqn:varianceS}
\end{eqnarray}
\vspace{-10mm}

\noindent
where $\phi(x)=\exp{-x^2/2}/\sqrt{2\pi}$. Then the asymptotic values of  {(\ref{eqn:meanS}) } are given by  
\vspace{-4mm}
\begin{eqnarray}\label{eqn:asympresult}
\lim_{N \rightarrow \infty} \mu_W-\bar{\mu} {\longrightarrow} 0\quad  \text{ and }\quad \lim_{N \rightarrow \infty} \bar{\sigma}^2_W-\bar{\sigma}^2 {\longrightarrow} 0. 
\end{eqnarray}
\vspace{-12mm}

	 \linespread{1.6}
\begin{proof}
 When $N$ grows large, by invoking the fact that 
 \begin{eqnarray}
\lim_{N \rightarrow \infty}  {\left(\sum_{n=1}^N x_n\right)}\Big/\sqrt{\sum_{n=1}^Nx^2_n}
\;\;{\longrightarrow} \;\;\infty,
 \end{eqnarray}
 where  $x_n=\eta_n\sqrt{\kappa_{g_n}\kappa_{h_n}}$, we can readily show that $\bar{Z} = -\bar{\mu}/\bar{\sigma} \rightarrow -\infty$, where $\bar{Z}$ is a parameter in (\ref{eqn:meanS}). Then, these results hold; $\phi(\bar{Z})\stackrel{{N\rightarrow \infty}}{\longrightarrow} 0,\quad \bar{Z}\phi(\bar{Z})\stackrel{{N\rightarrow \infty}}{\longrightarrow} 0,\quad \text{and}\quad 
 \xi\stackrel{{N\rightarrow \infty}}{\longrightarrow} 1.$ 
  By substituting these asymptotic values into  (\ref{eqn:meanS}), the desired result can be derived as shown in  (\ref{eqn:asympresult}).
\end{proof}

	 \linespread{0}
\textit{\textbf{Remark 3:}} The IRSs are meant to be  installed on  walls, ceilings, building facades, and advertisement panels \cite{Renzo2020}. They are comprised of low profiles with conformal geometry and light weight, and thus, they are cost effective \cite{Wu2020}. Thus,  IRSs with moderate to large number of reflective elements ($N$) are  a   feasible assumption in terms of   space requirements and cost. 

\begin{figure*}[!t]\centering \vspace{-10mm}
	\subfloat[ CDF for $N=32$]{\hspace{-0mm}
		\includegraphics[width=0.4\textwidth]{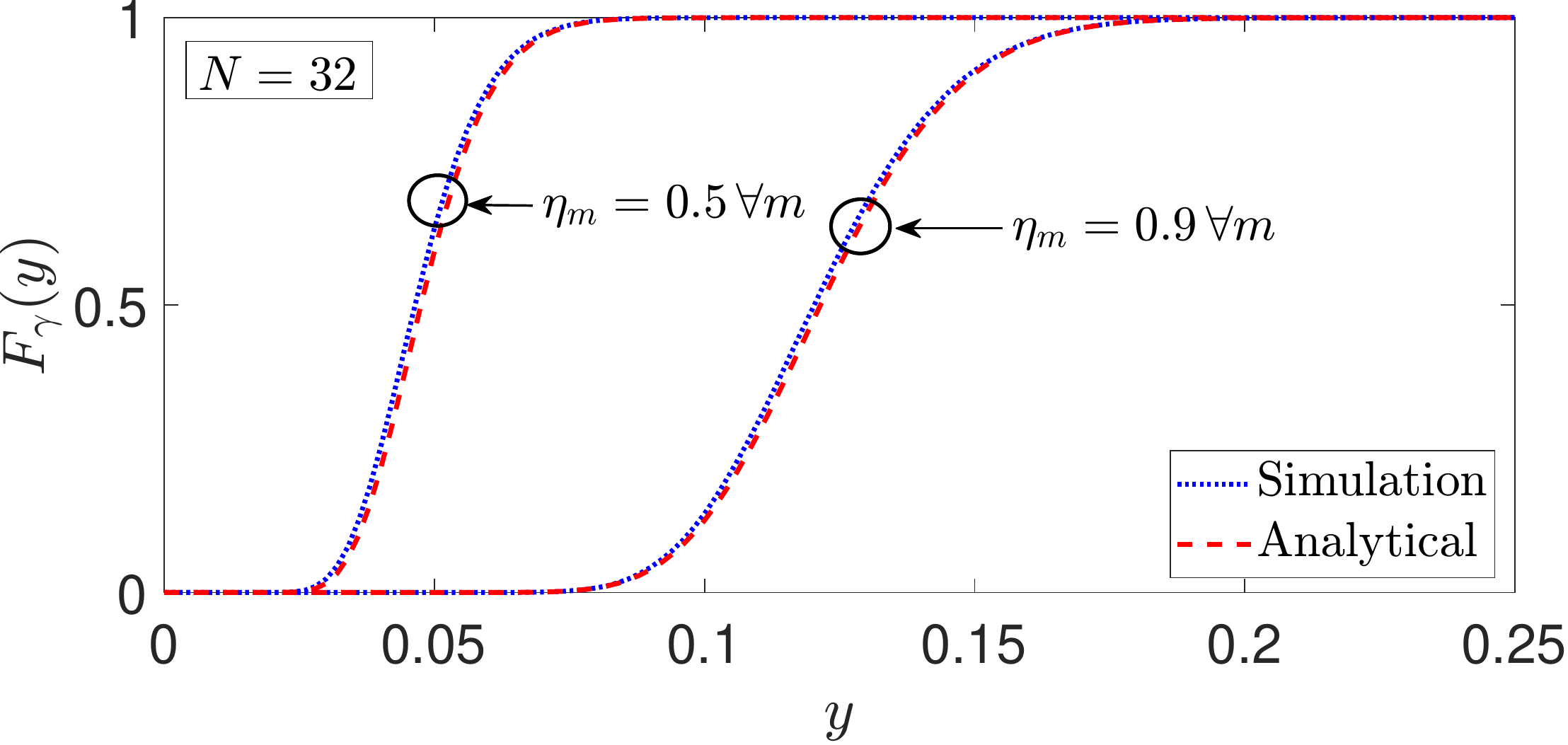}   \label{fig:CDF_N32} }
	\subfloat[ CDF for $N=256$]{\hspace{-0mm}
		\includegraphics[width=0.4\textwidth]{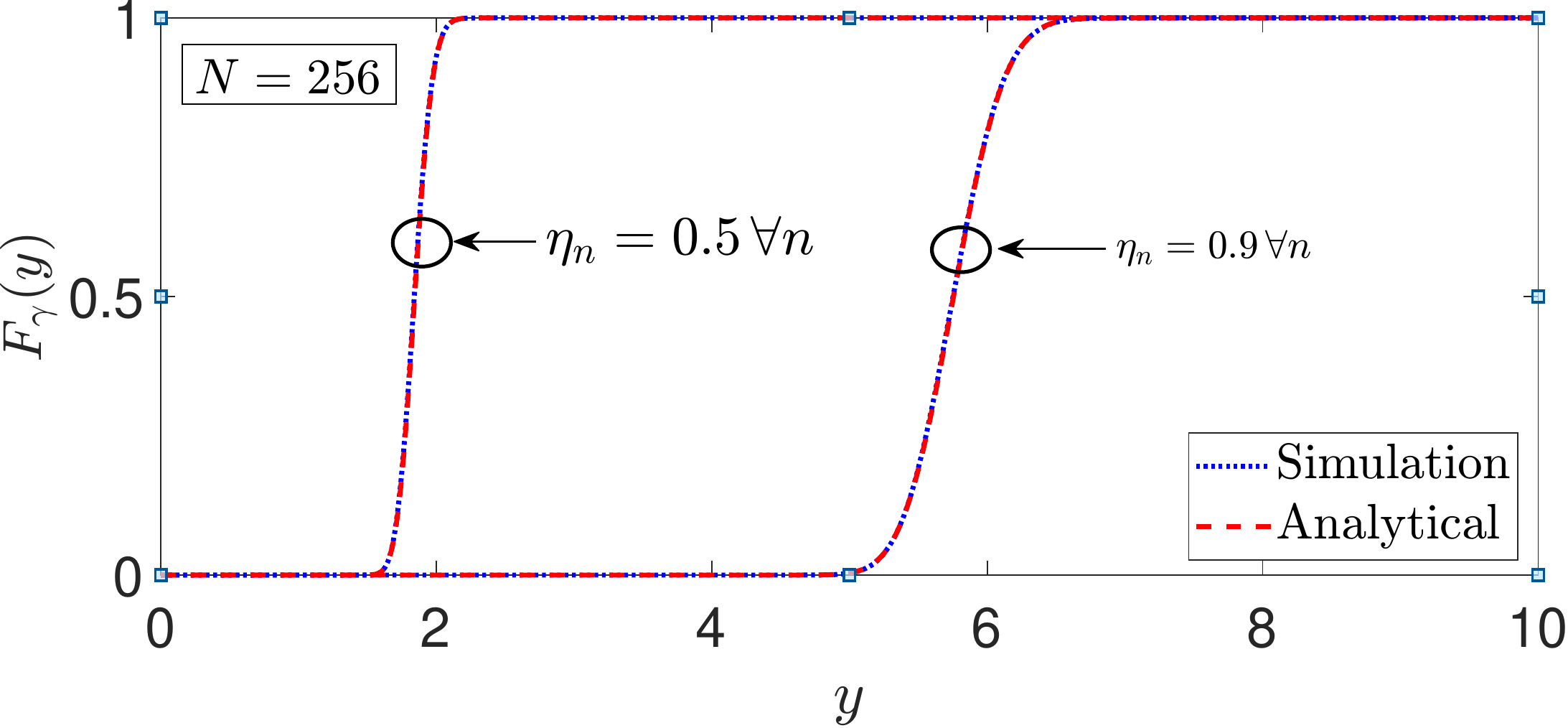}
		\label{fig:CDF_N256}  }\vspace{-2mm}
	\caption{The exact  CDF of $\tilde{\gamma}^*$ via Monte-Carlo simulations and  its analytical approximation $F_{\gamma}(y)$ in (\ref{eqn:cdf_SNR}).}\vspace{-10mm}
	\label{fig:CDF_SNR} 
\end{figure*}
By using {Lemma 2}, a tight approximation for the optimal received SNR   $(\tilde{\gamma}^*)$ is derived as

\vspace{-10mm}
\begin{eqnarray}\label{eqn:snr_opt_rwt}
\tilde{\gamma}^*\approx \gamma = \bar{\gamma} \left|\bar{v}+ W  \right|^2,
\end{eqnarray} 
\vspace{-10mm}

\noindent
where $W$ is our proposed tight approximation to $\tilde W$ in (\ref{eqn:W_tilde}).
Next, the CDF of $ \gamma$ is  derived as  

\vspace{-1mm}	
	\begin{small}
	\begin{eqnarray}\label{eqn:cdf_SNR}
	F_{\gamma}(y)=
	\begin{cases}
	\displaystyle 2\lambda\sqrt{a} \bar{\sigma}^2\sum_{k=0}^{\tilde m_v -k} \binom{\tilde m_v-k}{k} (-1)^{\tilde m_v-k}\left[\mathcal J\left(k,\frac{\bar{\mu}}{2\sqrt{a}\bar{\sigma}^2}\right)-\mathcal J\left(k,\frac{\bar{\mu}-\sqrt{y/\bar{\gamma}}}{2\sqrt{a}\bar{\sigma}^2}\right)\right], & 0<y\leq \bar{\gamma}\bar{\mu}^2\\
	1-\displaystyle 2\lambda\sqrt{a} \bar{\sigma}^2\sum_{k=0}^{\tilde m_v -k} \binom{\tilde m_v-k}{k}
	\left[{{[(-1)^k+1]}\Gamma\left(\frac{k+1}{2}\right)}\mathcal I\left(k,\frac{\sqrt{y/\bar{\gamma}}-\bar{\mu}}{2\sqrt{a}\bar{\sigma}^2}\right)
	\right.\\
	\left.\qquad \qquad \qquad \qquad \qquad \qquad  \qquad+\mathcal J\left(k,\frac{\sqrt{y/\bar{\gamma}}-\bar{\mu}}{2\sqrt{a}\bar{\sigma}^2}\right)\right], 
	&y> \bar{\gamma}\bar{\mu}^2,
	\end{cases}	
	\end{eqnarray}	
\end{small} 
\vspace{-4mm}
	
\noindent and $0$ otherwise. In (\ref{eqn:cdf_SNR}), $\tilde m_v$, $a, \lambda$ and  can be defined as 

\vspace{-10mm}
		\begin{eqnarray}
		\tilde m_v&=& 2m_v-1,\;\; a= {m_v}/{\kappa_v} + 1/(2\bar{\sigma}^2)  \label{eqn:def_a}\quad
		\text{ and }\quad 
		\lambda = {m_v^{m_v} \xi }\big/\left({ \Gamma(m_v) \kappa_v^{m_v} a^{m_v}\sqrt{2\pi \bar{\sigma}^2}}\right), \label{eqn:def_lambda}	
		\end{eqnarray}  		 
	\vspace{-10mm}
	
	\noindent
	where  $\bar{\mu}$ and $\bar{\sigma}^2$ are defined in {\eqref{eqn:meanS}.}
In (\ref{eqn:cdf_SNR}),  $\mathcal J\left(k,x\right)$ for $x\geq 0$ and $\mathcal I\left(k,x\right)$  can be defined as 

\vspace{-8mm}
	\begin{eqnarray}
	\mathcal J\left(k, x\right)  &=&\begin{cases}
	\mathcal J_o\left(k, x\right),  &\text{for} \quad k= 2t+1, t\in \mathbb Z,  \\
	\mathcal J_e\left(k, x\right),  &\text{for} \quad k= 2t, t\in \mathbb Z\cup \{0\},
	\end{cases}\label{eqn:Jkx}\\
	\mathcal I\left(k, x\right)&=&
	\frac{1}{2}\Gamma\left(\frac{k+1}{2}\right)+\frac{1}{2}\left(-1\right)^k\gamma\left(\frac{k+1}{2}, x^2\right)\label{eqn:Ikx} \text{ for } x<0,
	\end{eqnarray} 
and $\mathcal I\left(k, x\right)=\Gamma\left(\frac{k+1}{2},x^2\right)/2$ for $x\geq 0$. In (\ref{eqn:Jkx}),  $\mathcal J_o\left(k, x\right)$ and $\mathcal J_e\left(k, x\right)$ can  be  defined as 
	\begin{subequations}
	\begin{eqnarray}
	\mathcal J_o\left(k,x\right) &=& \frac{(\delta_o-1)!}{2}\sum_{i=0}^{\delta_e-1}\frac{1}{i!({2\bar{\sigma}^2 a})^{\delta_e+i}}\Gamma\left(m_v+i-\frac{k}{2},{2\bar{\sigma}^2 a}x^2\right),
	\label{eqn:I_mu_odd}
	\\
	\mathcal J_e\left(k,x\right) &=&
	\sum_{j=0}^{\delta_e-1}\frac{(\delta_e-1)!}{2(\Delta)^{\delta_e-j}j!}
	\left[x^{2j}\exp{-\Delta x^2}\Gamma\left(k',x^2\right)-\frac{1}{({2\bar{\sigma}^2 a})^{k'+j}}\Gamma\left(k'+j,{2\bar{\sigma}^2 a}x^2\right)\right],
	\label{eqn:I_mu_even1}
	\end{eqnarray}	
\end{subequations}	
where $\delta_o={(k+1)}/{2}$, $\delta_e=m_v-k/2$, $k'=({k+1})/{2}$, and $\Delta={2\bar{\sigma}^2 a} - 1$.

 The PDF of $\gamma$ is given by  $f_{\gamma}(y)=f_{R}(\sqrt{y/\bar{\gamma}})/(2\sqrt{y\bar{\gamma}})$, where $f_{R}(x)$ is defined in (\ref{eqn:pdf_R}).

	\begin{proof}
	See Appendix \ref{app:AppendixBB}.
\end{proof}

\vspace{-3mm}
\textit{\textbf{Remark 4:}} The tightness of our approximated CDF of the optimal  SNR (\ref{eqn:cdf_SNR}) derived via  (\ref{eqn:snr_opt_rwt}) is validated in Fig. \ref{fig:CDF_SNR}. Here,  we plot our  analytical CDF  (\ref{eqn:cdf_SNR}) for different $N$ and $\eta_n$, and it is compared against  the exact CDF of  the optimal   SNR   $(\tilde{\gamma}^*)$ via  Monte-Carlo simulations.  Fig. \ref{fig:CDF_SNR}   reveals that  our analytical CDF is   accurate   even for a moderately large number of IRS elements $(N=32)$, and this tightness significantly improves for the high  $N$ regime ($N=256$). Thus, this observation verifies that  our approximated yet   tractable statistical characterization  is accurate.   

%
\subsection{Outage probability }\label{sec:outage_prob}

The  SNR outage probability is defined as the probability that the instantaneous SNR $(\tilde{\gamma})$ falls bellow a threshold SNR\footnote{The rate outage probability can be defined as ${Pr}\left(\mathcal R = \mathrm{log}_2(1+\gamma)\leq \mathcal R_{th}\right)$, where  $\mathcal R$ and $\mathcal R_{th}$ are the achievable rate and  a  threshold, respectively. Thus,  it is readily related to the SNR outage probability; ${Pr}\left(\gamma\leq  2^{\mathcal R_{th}}-1\right) $.}   $(\gamma_{th})$. By using   (\ref{eqn:snr_opt_rwt}), a tight approximation to  the outage probability for the    moderately large $N$ regime  can be  written  as 

\vspace{-10mm}
\begin{eqnarray}\label{eqn:out_prob}
P_{out} = P_{r}\left(\tilde\gamma^* \leq \gamma_{th}\right) \approx F_{\gamma}(\gamma_{th}),
\end{eqnarray}  
\vspace{-10mm}

\noindent where $F_{\gamma}(\cdot)$ is given in \eqref{eqn:cdf_SNR}.

\textbf{\textit{Remark 5:}} To obtain   insights, we investigate the asymptotic outage probability in high SNR regime. The analytical behavior of outage probability (\ref{eqn:out_prob}) in high SNR regime can be  given by 

\vspace{-10mm}
	\begin{eqnarray}\label{eqn:Asymp_outage}
\lim_{\bar \gamma \rightarrow \infty}	 P_{out} = 	P^\infty_{out}\approx (O_c\bar{\gamma})^{-G_d}  +\mathcal{O}\left(\bar{\gamma}^{-(G_d+1)}\right),
	\end{eqnarray}
	\vspace{-10mm}
	
	\noindent
 where $G_d$ is the achievable diversity order, and $O_c$ is a measure of the array  gain \cite{Wang2003a}. By deriving a single-polynomial approximation for  $P_{out}$, the asymptotic outage probability can be derived as presented in  Theorem 1.
	
	\vspace{3mm}
	\textit{Theorem 1:} \textit{For the case of  $m_h\neq m_g$, when $\bar\gamma\rightarrow \infty$, the asymptotic outage probability  can be derived as 
		\begin{eqnarray}\label{eqn:Poutasym}
		P^\infty_{out}=\Omega_{op} \left({\gamma_{th}}\big/{\bar{\gamma}}\right)^{G_d}+\mathcal{O}\left(\left({\gamma_{th}}\big/{\bar{\gamma}}\right)^{(G_d+1)}\right),
		\end{eqnarray}
		where the diversity order $G_d$ is    given by 
		\begin{eqnarray}\label{eqn:diversity}
		G_d=m_v+\min(m_g,m_h)N.
		\end{eqnarray}	 
		In (\ref{eqn:Poutasym}), $\Omega_{op}$ is given by 
	%
		\begin{eqnarray}\label{eqn:omegaop}
		\!\!\!\!\!\!\!\!\!
		\Omega_{op}=\frac{2m_v^{m_v}\Gamma(2m_v)}{\Gamma(m_v)\kappa_v^{m_v}} \left[\frac{4(m_am_b)^{m_c} \Gamma(2m_c+\frac{1}{2})\Gamma(2m_b-2m_a-1)}{\Gamma(m_a)\Gamma(m_b)\Gamma(m_b-m_a+\frac{1}{2})\Gamma(2m_b)}\right]^N\left(\prod_{n=1}^N\frac{1}{\sqrt{\eta^2_n\kappa_{a}\kappa_b}}\right)^{m_c},
		\end{eqnarray}
		\noindent where $m_a=\min(m_g, m_h)$,  $m_b=\max(m_g,m_h)$, and $m_c=({m_a+m_b})/{2}$. Further,  $\kappa_{a}$ and $\kappa_{b}$ are scaling parameters of the corresponding channels. Moreover,  the asymptotic array gain   ($O_c$) in (\ref{eqn:Asymp_outage}) is given by $O_c = \gamma^{-1}_{th} \Omega^{-1/G_p}_{op}$.
				 When $m_g=m_h=m$, the diversity order reduces (\ref{eqn:diversity}) to $G_d=m_v+mN$. This confirms that the achievable diversity order can be linearly increased with the number of passive IRS elements without employing any additional active RF chains.} 
	\begin{proof}
		See Appendix  \ref{app:AppendixCC}.
	\end{proof}

\vspace{-3mm}
	\textit{\textbf{{Remark 6:}}} Our asymptotic outage analysis reveals that the  diversity order is {an} affine/linear function of the number of passive reflective elements ($N$). It is noteworthy to mention that this diversity gain is achieved without employing active RF chains at the IRS and  in  the presence of single-antenna $S$-$D$ pair. The diversity order of the direct transmission between $S$  and $D$ is $G_d=m_v$, where $m_v$ is the Nakagami-$m$ parameter of the direct channel. 
	Thus, the IRS-aided system achieves an additional diversity gain of $\min(m_g,m_h)N$, which is directly proportional to {$N$}. Thus, this diversity gain is attainable by recycling the existing EM waves via passive reflections without generating new EM waves from active RF chains. This diversity gain directly translates into reliability  improvements in terms of  outage probability and average SER.
	Thus, the system performance  can be  boosted   via  constructive signal combining at $D$  by virtue of intelligently controllable phases-shifts of passive  reflective elements at the IRS. 
	

\vspace{-2mm}
\subsection{Average achievable rate }\label{sec:achvble_rate}
The average achievable rate   can be defined as 

\vspace{-10mm}
	\begin{eqnarray}\label{eqn:avg_rate}
	\mathcal{R} = \E{\log[2]{1+\tilde{\gamma}^*}}\approx\E{\log[2]{1+\gamma}}, 
	\end{eqnarray}   
	where $\gamma$ is given in \eqref{eqn:snr_opt_rwt}. Since the exact derivation of the expectation   in \eqref{eqn:avg_rate} is mathematically involved, we derive tight upper and lower bounds by invoking Jensen’s inequality  as $\mathcal{R}_{lb} \leq \mathcal{R} \leq \mathcal{R}_{ub}$ \cite{Zhang2014}, where $\mathcal{R}_{lb}$ and $\mathcal{R}_{ub}$ are defined as
		\begin{eqnarray} 
		\mathcal{R}_{lb} &=& \log[2]{1+ \left(\E{1/\gamma}\right)^{-1}}  \label{eqn:rate_lb}\quad \text{and}\quad 
		\mathcal{R}_{ub} = \log[2]{1+ \E{\gamma}}. \label{eqn:rate_ub}	
		\end{eqnarray}  		 
	
	\textit{Theorem 2:} \textit{   
		By evaluating the expectations   in (\ref{eqn:rate_lb}), the tight achievable rate lower and upper bounds can be derived in closed-form as }
	\begin{eqnarray} \label{eqn:rate_lb_sub}
\!\!\!\!\!\!\mathcal{R}_{lb} = \log[2]{1+ \frac{\bar{\gamma} \left(\kappa_v + 2 \mu_{W}{\Gamma\left(m_v + 1/2\right)}\sqrt{{\kappa_v}/{m_v}}/{\Gamma\left(m_v\right)}  +\mu^2_W +\sigma_{W}^2 \right)^3}
	{\frac{\xi}{2\sqrt{\pi}}\left[\sum\limits_{k=0}^{4} \binom{4}{k} \frac{\Gamma\left(m_v+k/2\right)}{\Gamma\left(m_v\right)} \left(\frac{\kappa_v}{m_v}\right)^{k/2}    \sum\limits_{i=0}^{4-k}  \binom{4-k}{i} \left( {2 \bar{\sigma}^2}\right)^{i/2}  \bar{\mu}^{4-k-i} \mathcal I\left(i, \frac{-\bar{\mu}}{\sqrt{2 \bar{\sigma}^2}}\right)\right] }} ,
\end{eqnarray}
\begin{eqnarray}\label{eqn:rate_ub_sub}
\mathcal{R}_{ub} =   \log[2]{1+ \bar{\gamma} \left(\kappa_v + 2 \mu_{W}{\Gamma\left(m_v + 1/2\right)}\sqrt{{\kappa_v}/{m_v}}/{\Gamma\left(m_v\right)}  +\mu^2_W +\sigma_{W}^2 \right)},
\end{eqnarray}
where $\mathcal I(k,x)$ is defined in (\ref{eqn:Ikx}).
	\begin{proof}
		See Appendix \ref{app:AppendixDD}.
	\end{proof}

\vspace{-4mm}
	\subsubsection{Asymptotic achievable rate  when $N\rightarrow \infty$}
	
	To obtain further insights,  we derive the asymptotic rate when the number of reflective elements grows without bound. As per our achievable rate analysis in Section \ref{sec:achvble_rate}, in the asymptotic $N$ regime, the source transmit power  $(p)$ can be scaled inversely proportional to the square of the number of IRS elements ($N$). Thus, by scaling the transmit power at $S$ as $E=p/N^2$, we show that the upper and lower bounds of the achievable rate converge to a common asymptotic value  when the number of IRS elements grows without bound as follows: 
	\begin{eqnarray}
\lim_{N\rightarrow \infty}	\mathcal R_{lb}=\mathcal R^{\infty}   \qquad \text{ and }\qquad   \lim_{N\rightarrow \infty} \mathcal R_{ub}=\mathcal R^{\infty} , 
	\end{eqnarray}
	where the asymptotic rate $\mathcal R^{\infty}$ in the limit of infinitely many IRS elements is given by
	\begin{eqnarray}\label{eqn:asymptoticrate}
	\mathcal R^{\infty}=\log[2]{1+\bar{\gamma}_E{\eta^2\kappa_g\kappa_hT(m_g,m_h,1/2)}\big/({m_gm_h})},
	\end{eqnarray}
	where $\bar{\gamma}_E=E/\sigma^2_n$. This result also confirms that our achievable rate lower and upper bounds in (\ref{eqn:rate_lb_sub}) and (\ref{eqn:rate_ub_sub}) are asymptotically exact. 
	\begin{proof}
	{See Appendix \ref{app:AppendixDD}}.
\end{proof}	
	

\subsection{Average symbol error rate (SER)}
The average SER is defined as the expectation of the conditional error probability $(P_{e|\tilde \gamma^*})$ over the distribution of $\tilde{\gamma}^*$ \cite{Proakis2007}. For  a wide-range of modulation schemes,   $P_{e|\tilde\gamma^*}$  is given by   $P_{e|\tilde{\gamma}^*}=\alpha\Q{\sqrt{\beta\tilde \gamma^*}}$, where $\alpha$ and $\beta$ are modulation   dependent   parameters \cite{Proakis2007}. 
For instance, $(\alpha, \beta)$ for binary shift-keying (BPSK) and $M$-ary quadrature amplitude modulation (QAM) can be defined as $(1,2)$ and $(4(\sqrt{M}-1)/\sqrt{M},3/(M-1))$, respectively \cite{Andreas2005}.
Then,   the average SER can be derived as 
$\bar{P}_e=\E{\alpha\Q{\sqrt{\beta\tilde \gamma^*}}}$. By using {(\ref{eqn:snr_opt_rwt})}, $\bar{P}_e$ can be tightly approximated as

\vspace{-8mm}
\begin{eqnarray}\label{eqn:PeXX}
\bar{P}_e\approx\E{\alpha\Q{\sqrt{\beta \bar{\gamma}}(\bar{v}+W)}}.
\end{eqnarray}

	\textit{Theorem 3:} \textit{An upper bound for $\bar{P}_e$  can be derived as}
	\begin{eqnarray}\label{eqn:avg_ber}
	\bar{P}_{e} &\leq&\frac{\alpha\xi m_v^{m_v}\e{-\frac{\bar{\mu}^2}{2\bar{\sigma}^2}}}{ \kappa_v^{m_v}\sqrt{2}\bar{\sigma}^2}\frac{\e{\bar{\mu}^2\left[2\bar{\sigma}^2+\frac{2\beta\bar{\gamma}\bar{\sigma}^4}{\cos^2(\vartheta_u)}\right]^{-1}}}{\left(\frac{m_v}{\kappa_v}+\frac{\beta\bar{\gamma}}{2\sin^2(\vartheta_u)}\right)^{m_v}\sqrt{\frac{1}{2\bar{\sigma}^2}+\frac{\beta\bar{\gamma}}{2\cos^2(\vartheta_u)}}}\mathcal Q\left(-\frac{\sqrt{2}\bar{\mu}}{1+\frac{\beta\bar{\gamma}\bar{\sigma}^2}{\cos^2(\vartheta_u)}}\right),
	\end{eqnarray}
	\textit{where $\vartheta_u$, is given by}
	\begin{eqnarray}\label{eqn:vartheta}
	\vartheta_u&=&\underset{0\leq \vartheta\leq \pi/2}{\mathrm{argmax}}\;\;
	\left[\bar{\mu}^2\left(2\bar{\sigma}^2+\frac{2\beta\bar{\gamma}\bar{\sigma}^4}{\cos^2(\vartheta)}\right)^{-1}
	-m_v\ln\left(\frac{m_v}{\kappa_v}+\frac{\beta\bar{\gamma}}{2\sin^2(\vartheta)}\right)
	\right.\nonumber\\
	&&
	\left.-\frac{1}{2}\ln\left(\frac{1}{2\bar{\sigma}^2}+\frac{\beta\bar{\gamma}}{2\cos^2(\vartheta)}\right)
	+\ln\left(\mathcal Q\left(-\frac{\sqrt{2}\bar{\mu}}{1+\frac{\beta\bar{\gamma}\bar{\sigma}^2}{\cos^2(\vartheta)}}\right)\right)
	\right].
	\end{eqnarray} 
	\begin{proof}
		See Appendix \ref{app:AppendixEE}.
\end{proof}

\vspace{-3mm}
	\textbf{\textit{Remark 7:}} In order to draw useful insights, we derive the asymptotic average SER in  the high SNR regime. Thereby, we show that the  achievable diversity order is same as that resulted from the asymptotic outage probability analysis in  (\ref{eqn:Poutasym}). Moreover, the coding gain is also quantified. Thus, the asymptotic average SER is presented in Theorem 3.
	
	\textit{Theorem 3:} \textit{When $m_h\neq m_g$, the asymptotic SER as $\bar\gamma\rightarrow \infty$ can be derived as}
	\begin{eqnarray}\label{eqn:asympSER}
	P^\infty_{e}= \left(G_c\bar{\gamma}\right)^{-G_d} +\mathcal{O}\left(\bar{\gamma}^{-(G_d+1)}\right),
	\end{eqnarray}
	\textit{where the diversity order ($G_d$) is same as  (\ref{eqn:diversity}); $		G_d=m_v+\min(m_g,m_h)N$, and the array/coding gain ($G_c$) is given by  $G_c=\beta(\alpha2^{G_d-1}\Omega_{op}\Gamma(G_d+1/2)/\sqrt{\pi})^{-1/G_d},$
		where $\Omega_{op}$ is defined in (\ref{eqn:omegaop}).}
		\begin{proof}
		See Appendix \ref{app:AppendixCC}.
	\end{proof}

\vspace{-4mm}
\section{Effects of Quantized Phase--shifts} 
\vspace{-2mm}

In the previous section, we assumed that the IRS elements are capable of providing continuous phase-shifts  as defined by $\theta^*_n\;\; \forall n$ in (\ref{eqn:optimal_theta}). However, in practice, it may be prohibitively complicated to  enable real-time continuous phase-shifts at the IRS elements due to the hardware limitations. 
In this context, practical IRS deployments may adopt  discrete phase-shifts via phase quantization. To investigate the detrimental impact of quantized phase-shifts, we assume that the IRS controller is only able to select a limited number of discrete quantized phases   for the $n$th IRS element as  

\vspace{-10mm}
\begin{eqnarray}\label{eqn:descrete}
\hat{\theta}^*_n=
2\pi\hat{q}/2^b,\;\; \forall n,
\end{eqnarray} 

\vspace{-4mm}

\noindent
where $b$ is the number of quantization bits, and $\hat{q}$ is defined as
\vspace{-10mm}

\begin{eqnarray}
\hat{q}&=&\underset{q\in\{0,	\pm 1,\cdots, 	\pm 2^{b-1}\}}{\mathrm{argmin}}\big|\theta^*_n-\pi{q}/2^{b-1}\big|,\label{eqn:quantized}
\end{eqnarray}
where $\theta^*_n$ is  the optimal phase-shift defined in (\ref{eqn:optimal_theta}). The error between the unquantized and quantized phase-shift can be defined as 
\vspace{-12mm}

\begin{eqnarray}
\epsilon_n={\theta}^*_n-\hat{\theta}^*_n,
\end{eqnarray}

\vspace{-4mm}

\noindent
which tends to be uniformly distributed for a  large number of  quantization levels, i.e., $\epsilon_n\sim\mathrm{Uniform}\left[\right.-\tau,\tau\left.\right)$ for $\tau=\pi/2^b$ \cite{Haykin2009}. Moreover, this pahse error $\epsilon_n$ becomes uncorrelated with the signal for an increasing number of quantization levels \cite{Haykin2009}.  When IRS elements invoke discrete phase-shifts as per (\ref{eqn:descrete}), the optimal SNR  $\tilde\gamma^*$ in (\ref{eqn:SNR_Optimal}) can be rewritten as 
\begin{eqnarray}\label{eqn:SNR_Optimalquntized}
\hat{\tilde\gamma}^* &=&   { \bar{\gamma}\left| \bar{v} + \displaystyle \sum_{n=1}^{N} \bar{g}_n \bar{h}_n \eta_n\e{j\epsilon_n} \right| ^2 }= \bar{\gamma}\left(\left(\bar{v}+\tilde{W}_R\right)^2+\tilde{W}^2_I\right),
\end{eqnarray}  
where $\tilde{W}_R=\sum_{n=1}^{N} \bar{g}_n \bar{h}_n \eta_n\cos({\epsilon_n})$ and $\tilde{W}_I=\sum_{n=1}^{N} \bar{g}_n \bar{h}_n \eta_n\sin({\epsilon_n})$ account for the real and imaginary parts of the reflected signal from the IRS. By invoking {Lemma 2}, the distribution of $\tilde{W}_R$ and $\tilde{W}_I$ can be approximated as lower-tail truncated  normal  distributions. Specifically, the PDF/CDF of ${W}_R$ and ${W}_I$ can be deduced from (\ref{eqn:pdf_S}) and (\ref{eqn:CDF_S}) by replacing $(\bar{\mu},\bar{\sigma}^2,\xi)$ with $(\bar{\mu}_R,\bar{\sigma}^2_R,\xi_R)$ and $(\bar{\mu}_I,\bar{\sigma}^2_I,\xi_I)$, respectively. Here, $\bar{\mu}_R,\bar{\mu}_I,\bar{\sigma}^2_R,$ and $\bar{\sigma}^2_I$ can be defined as 
\begin{eqnarray}
\bar{\mu}_{R}&=&\bar{\mu}\sin(\tau)/\tau \qquad \text{and} \qquad     \bar{\mu}_{I}=0,\\
\bar{\sigma}^2_{R}&=&\left[\frac{\sin(2\tau)}{4\tau}+\frac{1}{2}\right]\left[\sum_{n=1}^N\eta_n\kappa_{g_n}\kappa_{h_n}\right]-\bar{\mu}^2_{R},
\text{ and }
\bar{\sigma}^2_{I}=\left[\frac{1}{2}-\frac{\sin(2\tau)}{4\tau}\right]\left[\sum_{n=1}^N\eta_n\kappa_{g_n}\kappa_{h_n}\right].
\end{eqnarray}
Moreover, $\xi_R=1/\mathcal Q(\bar{Z}_R)$, where $\bar{Z}_R=-\bar{\mu}_{R}/\bar{\sigma}_{R}$ and $\xi_I=1/\mathcal Q(0)=2$. By using (\ref{eqn:SNR_Optimalquntized}) and {(\ref{eqn:rate_lb}),} the lower and upper bounds for the  achievable rate with quantized phase-shifts are given by\footnote{The derivation follows steps similar to those in Appendix \ref{app:AppendixDD}, and hence, the proofs are omitted for the sake of brevity. }
	\begin{subequations}
	\begin{eqnarray} \label{eqn:rate_lb_subqyantized}
	\hat{\mathcal{R}}_{lb} = \log[2]{1+ \frac{\bar{\gamma} \left(\kappa_v + 2 \mu_{W_R}{\Gamma\left(m_v + 1/2\right)}\sqrt{{\kappa_v}/{m_v}}/{\Gamma\left(m_v\right)}  +\mu^2_{W_R} +\sigma_{W_R}^2 +\sigma_{W_I}^2 \right)^3}
		{\frac{(m_v+1)}{m_v}\kappa_v^2+4\bar{\mu}_R\frac{\Gamma(m_v+3/2)}{\Gamma(m_v)}\left[\frac{\kappa_v}{m_v}\right]^{3/2}+2(\bar{\mu}^2_R+\bar{\sigma}^2_R)\sigma^2_I+\sum_{k=1}^4 \mathrm I_k }} ,\qquad
	\end{eqnarray}
	\begin{eqnarray}\label{eqn:rate_ub_subqyantized}
	\hat{\mathcal{R}}_{ub} =   \log[2]{1+ \bar{\gamma} \left(\kappa_v + 2 \mu_{W_R}{\Gamma\left(m_v + 1/2\right)}\sqrt{{\kappa_v}/{m_v}}/{\Gamma\left(m_v\right)}  +\mu^2_{W_R} +\sigma_{W_R}^2 +\sigma_{W_I}^2 \right)},\qquad
	\end{eqnarray}
\end{subequations}
where $\mathrm I_k$ for $k\in\{1, \cdots,4\}$ are given by
\begin{subequations}
	\begin{eqnarray}
	\mathrm I_1&=&\frac{\xi_R}{2\sqrt{\pi}}\sum_{i=0}^4\binom{4}{i} (2\bar{\sigma}^2_R)^{i/2}\mathcal I \left(i,-\frac{\bar{\mu}_R}{\sqrt{2\bar{\sigma}^2_R}}\right),\label{eqn:I1}\quad
	\mathrm I_2=\frac{\xi_I}{2\sqrt{\pi}}\sum_{i=0}^4\binom{4}{i} (2\bar{\sigma}^2_I)^{i/2}\mathcal I \left(i,0\right),\label{eqn:I22}\\
	\mathrm I_3&=&\frac{4\Gamma(m_v+1/2)}{\Gamma(m_v)}\left[\frac{\kappa_v}{m_v}\right]^{1/2}
	\left[\frac{\xi_R}{2\sqrt{\pi}}\sum_{i=0}^3\binom{3}{i} (2\bar{\sigma}^2_R)^{i/2}\bar{\mu}^{3-i}_R\mathcal I \left(i,-\frac{\bar{\mu}_R}{\sqrt{2\bar{\sigma}^2_R}}\right)+\bar{\mu}_R\bar{\sigma}^2_I\right],\label{eqn:I3}\\
	\mathrm I_4&=&2\kappa_v\left(3(\bar{\sigma}^2_R+\bar{\mu}^2_R)+\bar \sigma^2_I\right).\label{eqn:I4}
	\end{eqnarray}
\end{subequations}
In (\ref{eqn:rate_lb_subqyantized}) and (\ref{eqn:rate_ub_subqyantized}), {${\mu}_{W_R}$, ${\sigma}^2_{W_I}$, and ${\sigma}^2_{W_R}$} are given by ${\mu}_{W_R}=\bar{\mu}_{R}+\bar{\sigma}_{R}\xi_R\phi(\bar{Z}_R)$, $\sigma^2_{W_I}=\bar{\sigma}^2\left(1\!-\!2/\pi\right)$, and $
\sigma^2_{W_R}=\bar{\sigma}^2_{R}\left(1+\bar{Z}_R\xi_R\phi(\bar{Z}_R)-(\xi\phi(\bar{Z}_R))^2\right)$,
where $\phi(\cdot)$ is defined in Lemma 2.

\textbf{\textit{Remark 9:}} When the  IRS consists of a large number of reflective elements and employs {quantized} phase-shifts, the rate bounds in (\ref{eqn:rate_lb_subqyantized}) and (\ref{eqn:rate_ub_subqyantized}) can be further simplified as follows: For moderately large $N$, by following steps similar to those used in Remark 2 and    by replacing $\bar{\mu}_{W_R}, \bar{\sigma}^2_{W_R},\bar{\sigma}^2_{W_I}$ and $\xi_R$ with $\bar{\mu}_{R}, \bar{\sigma}^2_{R},\bar{\sigma}^2_{I}$ and $1$, respectively,  in  (\ref{eqn:rate_lb_subqyantized}) and (\ref{eqn:rate_ub_subqyantized}), the  lower and upper bounds for the achievable rate with quantized phase-shifts can be deduced  as 
\begin{eqnarray} \label{eqn:rate_lb_subqyantized2}
\hat{\mathcal{R}}_{lb} = \log[2]{1+ \frac{\bar{\gamma} \left(\kappa_v + 2 \bar\mu_{R}{\Gamma\left(m_v + 1/2\right)}\sqrt{{\kappa_v}/{m_v}}/{\Gamma\left(m_v\right)}  +\bar{\mu}^2_{R} +\bar{\sigma}_{R}^2 +\bar{\sigma}_{I}^2 \right)^3}
	{\frac{(m_v+1)}{m_v}\kappa_v^2+2(\bar{\mu}^2_R+\bar{\sigma}^2_R)\sigma^2_I+2\kappa_v\left(3(\bar{\sigma}^2_R+\bar{\mu}^2_R)+\bar \sigma^2_I\right)+\sum_{k=1}^4 \bar{\mathrm I}_k }} ,\qquad
\end{eqnarray}
\vspace{-6mm}
	\begin{eqnarray}\label{eqn:rate_ub_subqyantized2}
\hat{\mathcal{R}}_{ub} =   \log[2]{1+ \bar{\gamma} \left(\kappa_v + 2 \bar{\mu}_{R}{\Gamma\left(m_v + 1/2\right)}\sqrt{{\kappa_v}/{m_v}}/{\Gamma\left(m_v\right)}  +\bar{\mu}^2_{R} +\bar{\sigma}_{R}^2 +\bar{\sigma}_{I}^2 \right)},\qquad
\end{eqnarray}
where 	$ \bar{\mathrm I}_k$ for $k=\{1,\cdots,4\}$ are given by $\bar{\mathrm I}_1 =4\bar{\mu}_R[\bar{\mu}^2_R+3\sigma^2_R+\sigma^2_I]{\Gamma(m_v+1/2)}\left({\kappa_v}/{m_v}\right)^{1/2}/{\Gamma(m_v)}$, $\bar{\mathrm I}_2 =4\bar{\mu}_R{\Gamma(m_v+3/2)}\left({\kappa_v}/{m_v}\right)^{3/2}/{\Gamma(m_v)}$, $\bar{\mathrm I}_3= 3 \sigma^2_I$, and $\bar{\mathrm I}_4=\bar{\mu}^4_R+6\bar{\mu}^2_R\bar\sigma^2_R+3\bar\sigma^2_R$.

\section{The impact of severity of fading and the underlying technical insights}

In this section, we discuss the impact of different {degrees} of severity of channel fading,  and thereby, we draw the underlying technical insights. {We} deduce some useful design insights from the derived performance metrics in Section  \ref{sec:performance_analysis}.  

\vspace{-2mm}
\subsection{Performance insights  in different fading environments}
\vspace{-2mm}
\subsubsection{All channels undergo Rayleigh fading}

When the system operates over Rayleigh fading, the   direct and reflected channel models can be deduced by setting the Nakagami-$m$ fading parameters as $m_v = m_g=m_h=1$. Specifically, Rayleigh fading models  {rich-scattering} and can be used to capture more severe fading environments than Nakagami-$m$ model for $m>1$ case.  In this context,   the outage probability, average SER, and  achievable rate bounds for Rayleigh fading can be deduced by letting  $m_v=m_g=m_h=1$ in (\ref{eqn:out_prob}), (\ref{eqn:avg_ber}), (\ref{eqn:rate_lb_sub}), and (\ref{eqn:rate_ub_sub}), respectively. Moreover, the achievable diversity order for the Rayleigh fading case is $G_d = 1 + N$.

\subsubsection{Direct channel undergoes severe fading}
 In a practical deployment, the IRS may be   placed at  specific  places such as on outer-walls of a high-rise building and windows in which the reflected channel may be accessible to a receiver. However,  it is not always guaranteed the availability  of a less severe  direct channel    between $S$ and  $D$. Thus, this direct channel is more likely to undergo much severe fading. This propagation condition can be modeled  by letting $m_v=1/2$ in (\ref{eqn:Nakagami_v}), which captures the most severe type of fading case \cite{Papoulis2002}. This choice  statistically  models the direct channel to be a  single-sided Gaussian fading. The outage probability for this asymmetric fading case can be  derived as 
 \vspace{-2mm}
\begin{eqnarray}\label{eqn:cdf_one_side_gaussian}
P^{(SG)}_{out} &\approx&  
1 - \frac{\lambda }{2}\bar{\sigma}^2\left[ \Gamma\left(\frac{1}{2},( ({\sqrt{\gamma_{th}/\bar{\gamma}}-\bar{\mu}})/{2\sqrt{a}\bar{\sigma}^2})^2\right)\exp{-\Delta ( ({\sqrt{\gamma_{th}/\bar{\gamma}}-\bar{\mu}})/{2\sqrt{a}\bar{\sigma}^2})^2} \right.\nonumber\\
&&\left.  
- \left(\Delta+1\right)^{-1/2}\Gamma\left(1/2, \left(\Delta+1\right)( ({\sqrt{\gamma_{th}/\bar{\gamma}}-\bar{\mu}})/{2\sqrt{a}\bar{\sigma}^2})^2\right) \right].\qquad
\end{eqnarray}
\vspace{-10mm}

\noindent
In deriving (\ref{eqn:cdf_one_side_gaussian}),  we assume that   the large-scale fading parameters {and}  the   reflection coefficient are the same for all reflective elements at the IRS;  i.e., $\kappa_{a_n}=\kappa_a$ for $a\in\{g,h\} $) and $\eta_n=\eta, \;\forall n$ for the sake of mathematical tractability.
The corresponding achievable rate {and average SER} bounds can be readily obtained by letting $m_v=1/2$ 	in (\ref{eqn:rate_lb_sub}), (\ref{eqn:rate_ub_sub}), and (\ref{eqn:avg_ber}), respectively.

\subsubsection{Direct channel is unavailable}
The direct channel between $S$ and $D$ may be fully unserviceable due to severe blockage effects. In this case, the end-to-end communication takes place solely via the IRS reflected channel. {Then,} the   outage probability, average SER, and   rate bounds can be derived by letting $\kappa_v\rightarrow 0$ (equivalently, $\zeta_v\rightarrow 0)$ in   (\ref{eqn:out_prob}), (\ref{eqn:avg_ber}), (\ref{eqn:rate_lb_sub}), and (\ref{eqn:rate_ub_sub}).

\vspace{-4mm}
\subsection{Impact of spatially-correlated fading}
\vspace{-3mm}

Once a large number of reflective elements is embedded with an IRS with a fixed area, the incident and reflected channels must be modeled to capture spatially correlated fading. To this end, 
the  correlation effects of $S$-to-IRS and IRS-to-$D$ channels are modeled by considering  IRS as a uniformly distributed 2D square array of reflective elements. Without-loss of generality, it is assumed that IRS is placed in the  far-field of $S$ and $D$. The angles-of-arrival (AoAs) and angles-of-departure  (AoDs) at the IRS are assumed to be randomly distributed  in azimuth and elevation planes as $\Omega_i\sim \mathcal{CN}(\omega_i,\nu^2_i)$ and $\Psi_i\sim \mathcal{CN}(\psi_i,\delta^2_i)$, respectively, where $i=$A and $i=$D are used to denote AoA and AoD.
Then, the spatial correlation   at the IRS can be modeled as $\mathbf R_i = \mathbf R_{i,az}\otimes\mathbf R_{i,el}$, where $\mathbf R_i$ for $i\in\{\text{A},\text{D}\}$  are the spatial correlation matrices at the IRS for incident and reflected signals. Here, $\mathbf R_{i,el}$ and $\mathbf R_{i,az}$ can be modeled as  \cite{Kudathanthirige2017}
\vspace{-2mm}
\begin{eqnarray}
\left[\mathbf R_{i,el}\right]_{x,y}&=&\exp{j\left({2\pi d_{el}}(y-x)\cos \psi_i\right)}
\exp{-\left(\frac{1}{2}\left(\delta_i{2\pi d_{el}}\right)^2(y-x)^2\sin^2 \psi_i\right)},\label{eqn:R_el}\\
\left[\mathbf R_{i,az}\right]_{r,t}&=& a^{-1/2}_1 \exp{-\left(\frac{a_2\cos^2 \omega_i}{2a_3}+\frac{(a_1{\nu}_i)^2\sin^2\omega_i}{2a_3}\right)}\exp{j\frac{a_1\cos \omega_i}{a_3}},\label{eqn:R_az}
\end{eqnarray}
where $a_1=(2\pi d_{az})(t-r)\sin {\psi_i}$, $a_2=\delta_i(2\pi d_{az})(t-r)\cos {\psi_i}$, and $a_3=a^2_2\nu^2_i\sin^2 \omega_i+1$. In (\ref{eqn:R_el}) and (\ref{eqn:R_az}), $d_{az}$ and $d_{el}$ are the  distances between the adjacent elements (measured in multiples of wavelengths)
in elevation and azimuth planes, respectively.  Under  spatially correlated fading, the $S$-to-IRS and IRS-to-$D$ channels can be modeled as
\vspace{-4mm}
\begin{eqnarray}\label{eqn:channel_with Corerelations}
\tilde{\mathbf g}^T= \mathbf g^T\mathbf R^{1/2}_{\text{D}} \qquad\text{ and }\qquad \tilde{\mathbf h}=\mathbf R^{1/2}_{\text{A}}\mathbf h,
\end{eqnarray}
\vspace{-10mm}

\noindent
 where ${\mathbf g}^T=[g_1\cdots, g_n\cdots, g_N]$ and $ {\mathbf  h}=[h_1\cdots, h_n\cdots, h_N]^T$. Here $h_n$ and $g_n$ are defined in Section \ref{sec:channel_model} and the resulting SNR can be written as 
\begin{eqnarray}\label{eqn:SNR2}
\tilde{\gamma} = { \bar{\gamma}\Big| \bar{v}\e{j\phi_v}+{\mathbf g}^T\mathbf R^{1/2}_{\text{D}}\mathbf \Theta \mathbf R^{1/2}_{\text{A}}{\mathbf h} \Big| ^2 },
\end{eqnarray} 
where $\mathbf \Theta$ is an $N\times N$ diagonal matrix whose $n$th diagonal element is given by $\left[\mathbf \Theta\right]_{n,n}=\eta_n \exp{j\theta_n}$, which denotes the reflection coefficient of the $n$th IRS element.

We observe that in the presence of spatially correlated fading, the IRS phase-shifts can be optimized 
by also considering the  phases introduced by correlation matrices in (\ref{eqn:R_el}) and (\ref{eqn:R_az}) in addition to the phases of the independently faded channels $\tilde{\mathbf g}^T$ and $\tilde{\mathbf h}$.
To this end, the phase-shift of each IR element, which maximizes the SNR, can be computed as
\vspace{-3mm}
\begin{eqnarray}
\theta^*_n&=& \phi_v-(\tilde{\phi}_{h_n}+\tilde{\phi}_{g_n}), \;\;\; \text{for } 1\leq n\leq N,\label{eqn:scheme2}
\end{eqnarray} 
where $\tilde{\phi}_{g_n}$ and $\tilde{\phi}_{h_n}$ are the phases of $ [\tilde{\mathbf g}^T ]_{1,n}$ and  $ [\tilde{\mathbf h} ]_{n,1}$, respectively, and they include the cumulative phases   $(\mathbf g^T,\mathbf h)$ and $(\mathbf R^{1/2}_{\text{D}},\mathbf R^{1/2}_{\text{A}})$ in (\ref{eqn:channel_with Corerelations}). Then, the optimal SNR is given by
\begin{eqnarray}
\tilde{\gamma}^*= \bar{\gamma}\Big| \bar{v}+\hat{\mathbf g}^T\boldsymbol {\eta} \hat{\mathbf h} \Big| ^2. \label{eqn:scheme21}
\end{eqnarray}  
where $\boldsymbol{\eta}$ is $N\times N$ diagonal matrix with $\eta_n$ as  the $n$th diagonal element. Moreover,   $\hat{\mathbf g}^T$ and $\hat{\mathbf h}$ denote the vectors with the moduli of the elements of ${\mathbf g}^T$ and ${\mathbf h}$, respectively.

  \vspace{-3mm}
 \section{Numerical results}\label{sec:Numerical_resilts}
 
   \vspace{-3mm}
%
   
   {
   
%
   
   In this section, our numerical results are presented to validate our analysis, investigate performance gains,  and   draw useful  design insights. 
   In our simulations, the path-loss is modeled as $\zeta_a\,[\text{dB}]=\zeta_0+10\upsilon \log{d}$, where $\zeta_a\in\{\zeta_v,\zeta_g,\zeta_h\} $,  $\zeta_0=42$\,dB is a reference path-loss,  $\upsilon=3.5$ is  the path-loss exponent, and $d$ is the distance in meters. 
   Unless otherwise specified, the amplitude attenuation coefficient is set to $\eta_n=0.9, \forall n\in\{1,\cdots, N\}$. 
   
%
\begin{figure*}[!t]\centering \vspace{-5mm}
	\subfloat[The average achievabe rate for different   $N$.]{\hspace{-0mm}
		\includegraphics[width=0.40\textwidth]{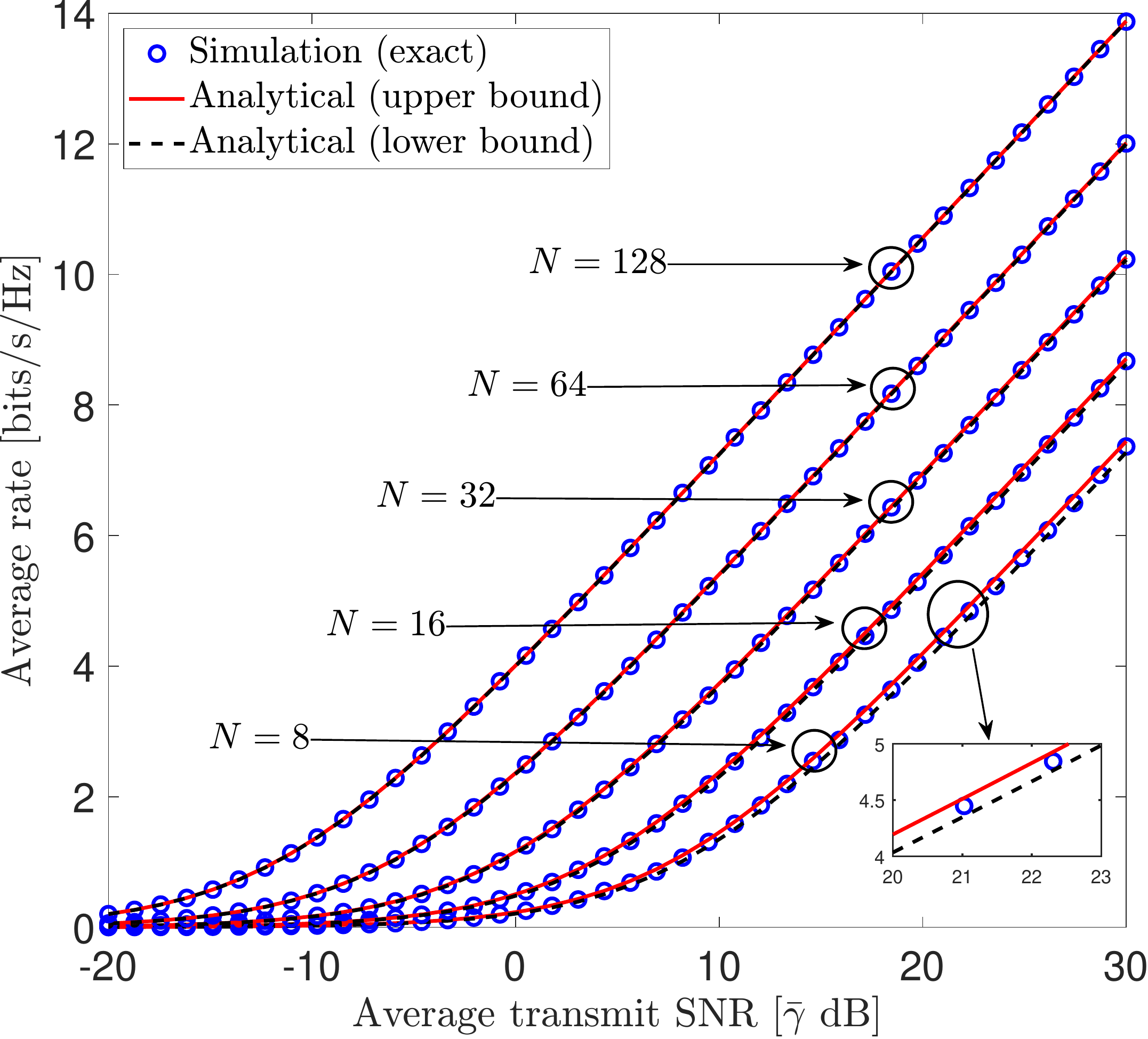}   \label{fig:rate}  }
	\subfloat[A comparison of the   rate for different disntances.]{\hspace{15mm}
		\includegraphics[width=0.40\textwidth]{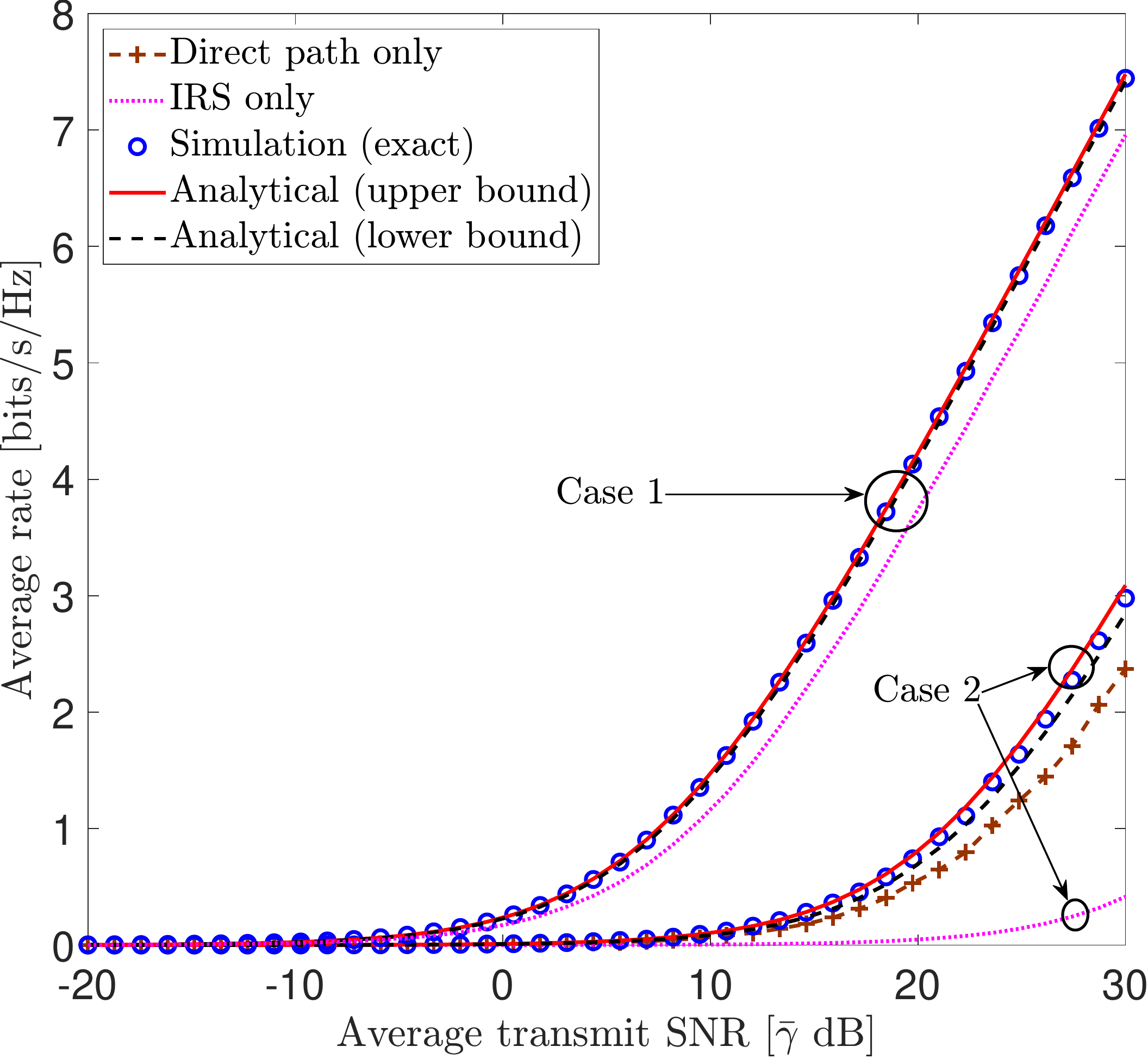}
		\label{fig:rate_comparison} }
	\caption{The average rate versus average transmit SNR for $(m_v,m_g,m_h)=(2,3,4)$ and $d_{SD}=100$\,m. In Fig. \ref{fig:rate}, $d_{SI}=d_{DI}=60$\,m.  In Fig. \ref{fig:rate_comparison}, the distances for Case-1 and Case-2 are set to  $d_{SI}=d_{DI}=60$\,m, and $d_{SI}=d_{DI}=140$\,m.}\vspace{-10mm}
	\label{fig:Rate} 
\end{figure*}
   
   
   In Fig. \ref{fig:rate}, we plot the average achievable rate   as a function of average transmit SNR  for different numbers of reflective elements at the IRS as $N=[8,16,32,64,128]$. The analytical curves for the lower and upper rate bounds   are plotted via  (\ref{eqn:rate_lb_sub}) and (\ref{eqn:rate_ub_sub}), respectively. The exact/optimal achievable rate is also plotted via Monte-Carlo simulations for comparison purposes. We observed from  Fig. \ref{fig:rate} that   our rate bounds are tight even for a small number of IRS elements such as $N=8$, and they converge to the exact simulation with increasing $N$ (see $N=128$ case in Fig. \ref{fig:rate}). Moreover,  higher the  number of reflective elements at IRS, higher   the achievable rate. For example, at an average SNR of $20$\,dB,  the average rate can be  increased by $1.9$\,bits/s/Hz and $3.1$\, bits/s/Hz, respectively, when the number of IRS elements is doubled and quadrupled  in comparison to  $N=16$ case.
   
%
%
%


Fig. \ref{fig:rate_comparison} provides a comparison of the average achievable rate for the IRS-assisted system with respect to a baseline direct transmission between $S$ and $D$. Moreover, the average rate for the IRS only case (without the direct channel) is also plotted. To highlight the performance gain of IRS, two system configurations in which   IRS is placed such that $d_{SD}\approx d_{SI}+ d_{ID}$ (Case-1)  and $d_{SD}\ll d_{SI}+ d_{ID}$ (Case-2) are considered. Fig. \ref{fig:rate_comparison} clearly shows that the cases, where  the direct transmission is aided by  IRS, {outperforms} all other cases  throughout the useful SNR regime regardless of the distance to the IRS. When the  direct channel is unserviceable, the IRS-aided system can provide  significant rate gains provided that the   IRS is placed nearer to $S$ and $D$ as in Case-1. However, when the IRS is placed further away from $S$ and $D$, the distance-dependent path-loss effects hinder   the rate performance gains of the IRS. Specifically,   at an average SNR of 25\,dB, Case-1 provides a rate gain of 4.492\,bits/s/Hz over the direct transmission. However, in Case-2, a rate gain of 0.486\,bits/s/Hz is achieved over the direct transmission at the same average SNR. Thus, the placements of IRS with respect to $S$ and $D$ has a significant impact on the achievable rate gains. This is primarily due to the fact that  beamforming gains attainable via  passive reflections at the IRS may not be able to overcome the larger path-losses   when the IRS is placed far away from $S$ and $D$.

\begin{figure}[!t]\centering
	\includegraphics[width=0.4 \textwidth]{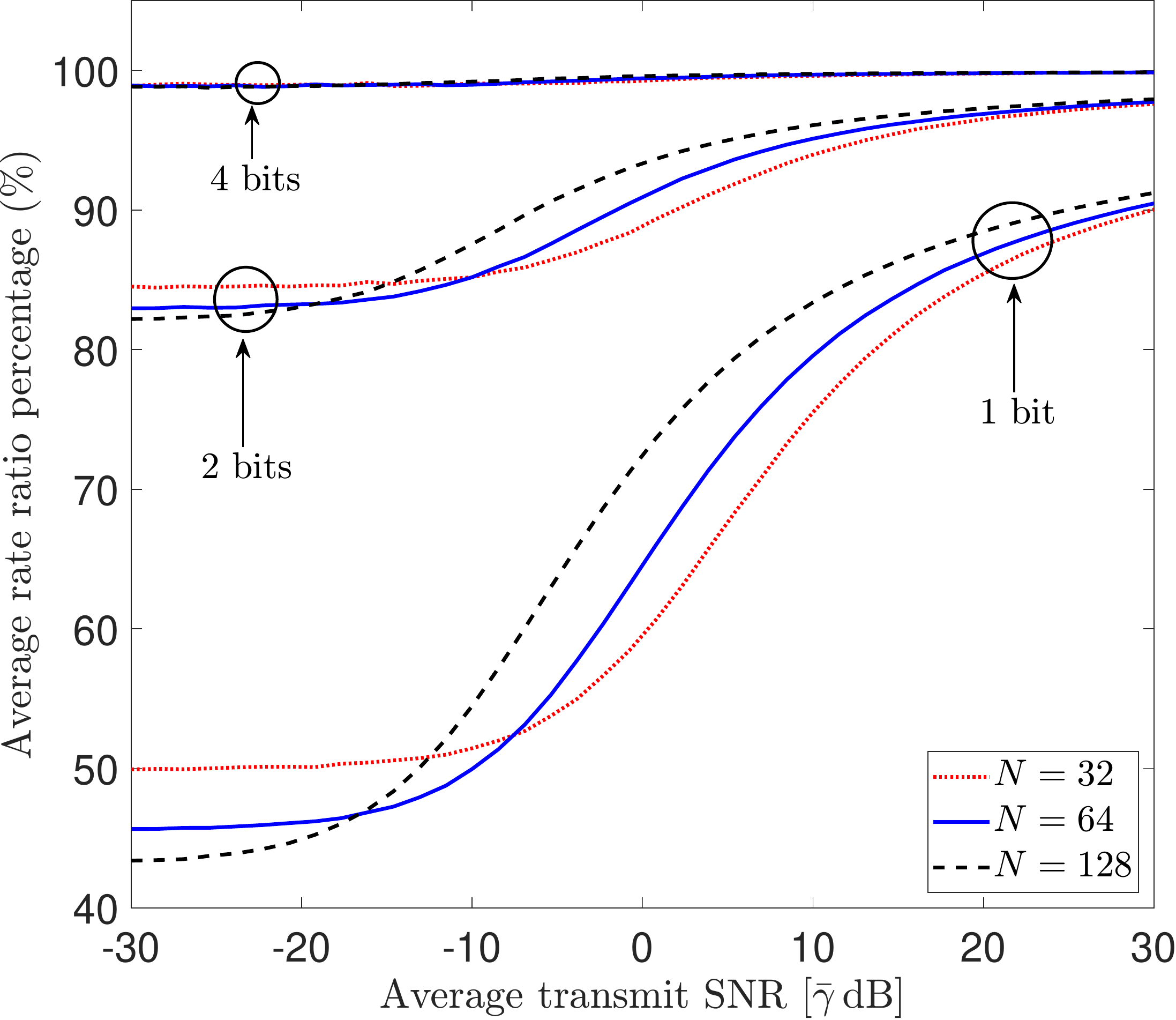}\vspace{-3mm}
	\caption{{The impact of phase-shift quantization  on the average achievable rate   different $N$ values. $\text{The average rate percentage} = \frac{\hat{\mathcal{R}}}{\mathcal{R}} \times 100\%$, where $\hat{\mathcal{R}}$ and $\mathcal{R}$ are the   average  achievable rates with and without phase-shift quantization errors, respectively. Here,  $(m_v,m_g,m_h)=(2,3,4)$, $d_{SD}=100$\,m, and $d_{SI}=d_{DI}=60$\,m.}}
	\label{fig:rate_ratio}\vspace{-10mm}
\end{figure}

Fig. \ref{fig:rate_ratio} is used to investigate the  impact of the number of quantization bits $(b)$ used for IRS phase-shift controlling, where the phase quantization error is randomly generated within $[-\pi/2^b,\pi/2^b]$.  The underlying detrimental impact is quantified by computing the average rate percentage, which is defined as  the percentage of the rate gain  that can be achieved by employing quantized phase-shifts at the IRS elements with respect to the benchmark system, where continuous phase-shifts are allowed with  no phase errors. In this context, the average rate percentage can be defined as follows: $\text{average rate percentage} = \frac{\hat{\mathcal{R}}}{\mathcal{R}} \times 100\%$.
where $\hat{\mathcal{R}}$ and $\mathcal{R}$ are the   average  achievable rates with and without phase-shift quantization errors, respectively. According to Fig. \ref{fig:rate_ratio}, when $b$ is increased, the effect of phase errors becomes negligible. In particular, with  $4$ bit quantization at all IRS elements, more than $98$\% of the average rate can be recovered with respect to the continuous phase-shifts. Moreover, the average rate percentage improves  with the transmit SNR. For example, more than $80$\%, $95$\%, and almost $100$\% of the rate gains can be recovered with $1$, $2$ and $4$ bit quantization, respectively, when the transmit SNR is maintained beyond $20\,$dB ($\bar{\gamma}\geq 20\,$dB). Moreover, in Fig. \ref{fig:rate_ratio}, the average rate percentage curves are plotted for $N=[ 32,\; 64,\; 128]$. While increasing number of IRS elements is beneficial in moderate-to-large SNR regime, it detrimentally affects the rate performance in low SNR regime. This is primarily due to  the adverse effects of  quantization errors becoming increasingly dominant with increasing $N$ in the  low SNR  regime.

\begin{figure}[!t]\centering
	\includegraphics[width=0.4 \textwidth]{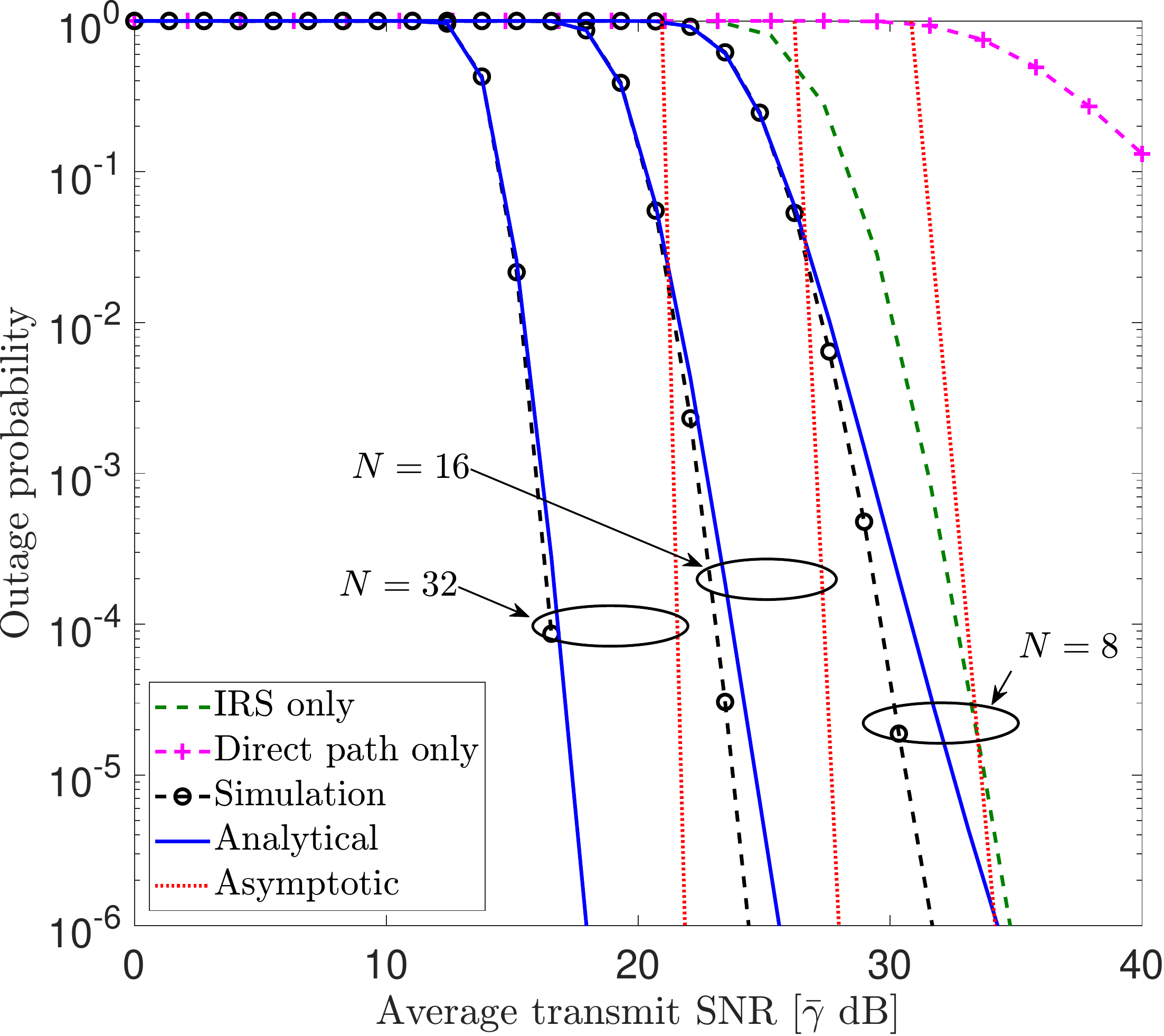}\vspace{0mm}
	\caption{The outage probability versus the average transmit SNR for different numbers of IRS elements with $\gamma_{th}=10$ dB, $(m_v,m_g,m_h)=(2,2,3)$, $d_{SD}=100$\,m, and $d_{SI}=d_{DI}=60$\,m.}
	\label{fig:outage}\vspace{-10mm}
\end{figure}

In Fig. \ref{fig:outage}, the outage probability of the IRS-aided system is investigated.  To this end,  our outage probability  (\ref{eqn:out_prob}) and the asymptotic outage in high SNR regime  (\ref{eqn:Poutasym}) are plotted
as a function of  the average transmit SNR  for $N=[8,\; 16,\; 32]$. Moreover, 
 Monte-Carlo simulations for the outage probability {are} presented. Tightness of our outage probability  analysis   improves with increasing $N$. The negative gradient of the asymptotic outage curves in Fig. \ref{fig:outage} can be computed to obtain the achievable diversity order. For instance, in the case of $N=16$, the negative gradient of the   asymptotic outage curve  is {$-(\log[10]{0.9512}-\log[10]{0.00001944})\times10/(26.21-27.59) \approx 34$}, which exactly agrees with our diversity order analysis in (\ref{eqn:diversity}); $G_d =m_v+\min(m_g,m_h)N = 2+ \min(2,3)\times 16= 34$. Moreover, the achievable diversity order can be boosted {266.6\%} by increasing the number of reflected elements from  $N=8$ to $N=32$.  Thus, this observation verifies our asymptotic performance analysis that the 
 diversity order can be drastically increased by increasing $N$ of the IRS.  It is noteworthy to mention that this diversity gain is achieved by merely employing passive IRS reflective elements without using costly active RF chains. 
  Moreover, in Fig. \ref{fig:outage}, the outage probability curves of the   direct channel (without the IRS) and sole IRS-aided   $(N=8)$ system (without the direct channel)  are plotted for  comparison purposes. These outage curves  clearly depicts that the passive reflections of the IRS  can indeed boost the outage performance.   For instance, the sole direct transmission cannot achieve an outage of $10^{-2}$ even with an average SNR of  {40}\,dB. However, when an IRS with $N=8$ is deployed without the direct channel, an average SNR of  {30}\,dB is sufficient to achieve the same outage probability.  

\begin{figure*}[!t]\centering \vspace{-5mm}
	\subfloat[The average BER for different system configurations.]{\hspace{-0mm}
		\includegraphics[width=0.42\textwidth]{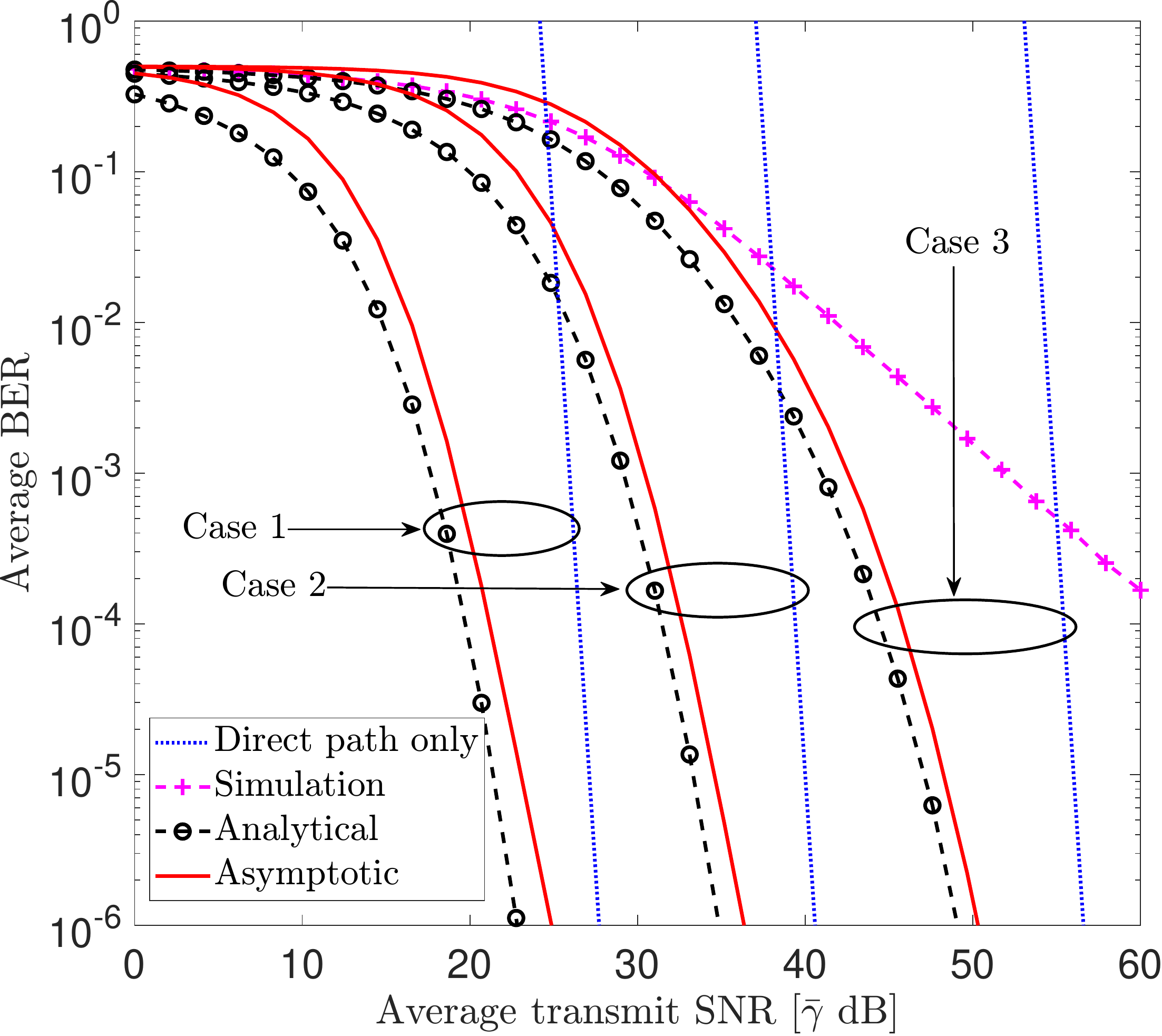}   \label{fig:SER} }
	\subfloat[The average BER versus the number of IRS elements.]{\hspace{15mm}
		\includegraphics[width=0.40\textwidth]{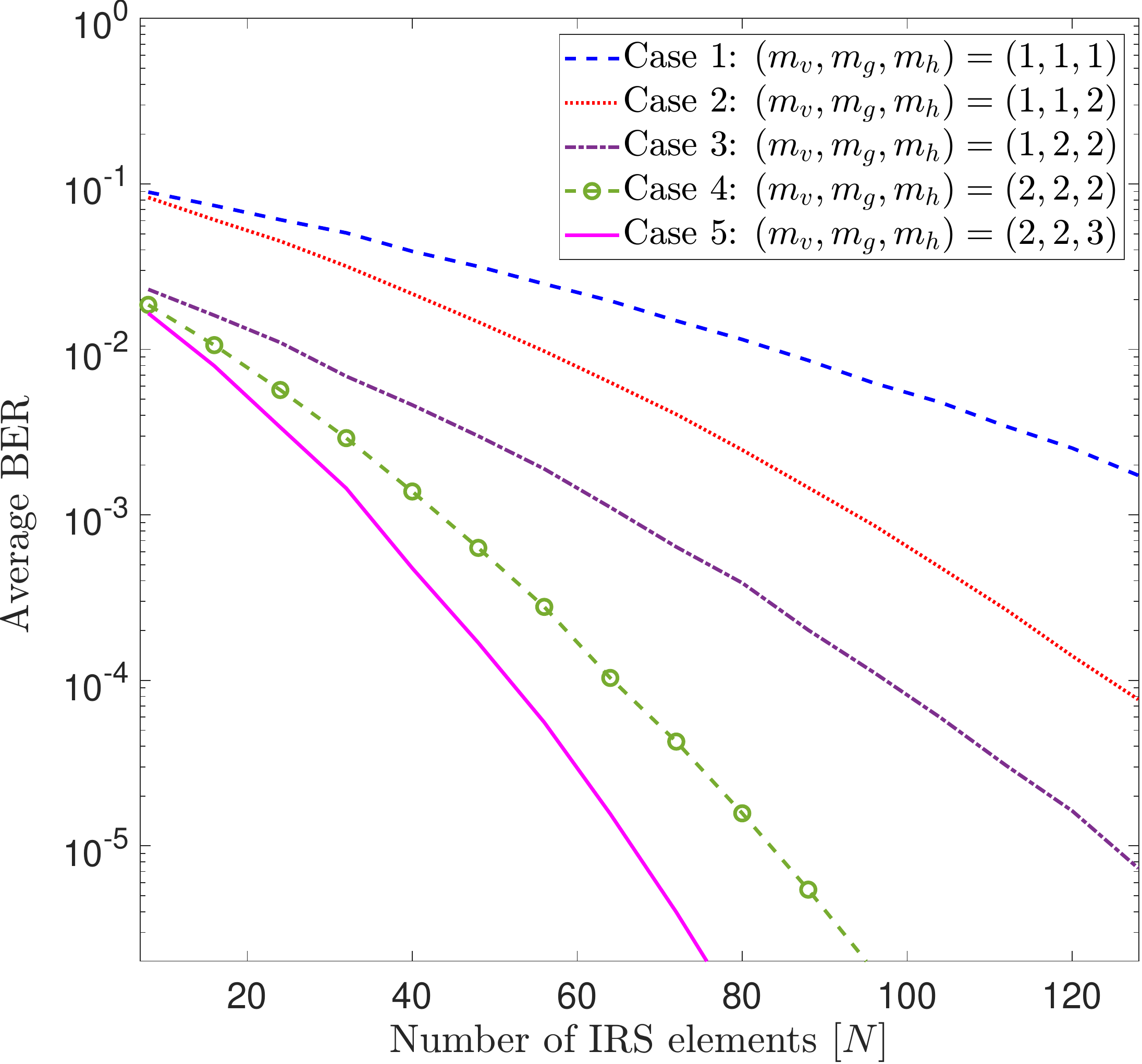}
		\label{fig:ber_comparison} }
	\caption{An average BER comparison for BPSK with $\alpha=1$, $\beta=2$ in (\ref{eqn:avg_ber}). For all cases, $d_{SD}=100$\,m. In \ref{fig:SER}, $(m_v,m_g,m_h)=(1,1,2)$, $N=16$,  the distances for Case-1 to Case-3 are set to  $d_{SI}=d_{DI}=51$\,m, $d_{SI}=d_{DI}=80$\,m, and $d_{SI}=d_{DI}=140$\,m, respectively. In Fig. \ref{fig:ber_comparison}, $d_{SI}=d_{DI}=140$\,m and $\bar{\gamma}=30$\,dB.}\vspace{-10mm}
	\label{fig:BER} 
\end{figure*}


{In  Fig. \ref{fig:SER},  the average bit error rate  (BER) of BPSK and the asymptotic BER analysis in high SNR regime are plotted and are compared with the direct transmission between $S$ and $D$. In Cases-1, 2 and 3,  the IRS is placed, respectively,  at a closer location to $S$ and $D$ such that  $d_{SD}\approx d_{SI}+ d_{ID}$ (Case-1), at a moderate distance according to $d_{SD} < d_{SI}+ d_{ID}$ (Case-2), and at a far-away location such that $d_{SD}\ll d_{SI}+ d_{ID}$ (Case-3). The  transmission distance of the direct channel is $d_{SD}$ at 100\,m. Since both direct and reflected signals are co-phased, the IRS-assisted system outperforms the    direct transmission in all three cases. For example, to obtain  an average BER of $10^{-2}$, the direct transmission needs an average SNR of $41.86$\,dB. However, when an IRS is placed at a far-away location (Case-3), this SNR requirement can be reduced by 4\,dB. Furthermore, when compared with the direct transmission, to achieve the same average BER of $10^{-2}$, the SNR requirement can be significantly lowered by around 10.5\,dB and 21.5\,dB when the IRS is placed at moderate (Case-2) and closer (Case-1) distances, respectively. Thus, these observations reveal that the average BER performance of IRS-assisted systems heavily depends on the placement of the IRS with respect to $S$ and $D$.    This is because, IRS only facilitates the end-to-end direct transmission by virtue of passive reflections, and if the transmission distances of the reflected channels are high, then the underlying large-scale fading losses cannot be compensated by passive beamforming gains provided by phase-shift control at the IRS.  The   negative gradient of the asymptotic BER curves in Fig. \ref{fig:SER} is $-(\log[10]{0.01}-\log[10]{0.0001})10/(38.27-39.47)\approx 17$.  This observation verifies our high SNR BER analysis in (\ref{eqn:diversity}) such that   $G_d =m_v+\min(m_g,m_h)N =1+\min(1,2)\times 16 =17$. It is also worth noting that the diversity orders for all three cases are the same because three average BER curves are plotted by varying the   distances and  keeping $N$, $m_v$, $m_g$, and $m_h$ fixed.   
}

{In Fig. \ref{fig:ber_comparison}, the effects of number of reflective elements ($N$) and  asymmetric channel fading conditions are investigated by plotting  the average BER of BPSK  as a function of $N$ by varying the severity of fading parameters $(m_v, m_g, m_h)$.   Fig. \ref{fig:ber_comparison}   reveals that the average BER  performance improves with higher $N$. For example, the numbers of IRS elements required to achieve an average BER of   $10^{-2}$ at $\bar{\gamma}=30$\,dB  are $84$, $56$, $26$, $16$, and $9$  for Case-1 to Case-5, respectively. Moreover, the average BER reduces with higher  $(m_v, m_g, m_h)$. This is because higher $(m_v, m_g, m_h)$, less severe the fading conditions as per Nakagami-$m$ fading model. Thus, this observation indicates that by placing  the IRS  such that the reflected channels undergo  milder fading conditions can improve the performance of IRS-aided systems. This insight is in-line with the original idea of coating building walls, windows and  hallways with IRS to minimize channel blockages/obstructions and thereby facilitate reliable wireless access.  }

}

\begin{figure}[!t]\centering
	\includegraphics[width=0.40 \textwidth]{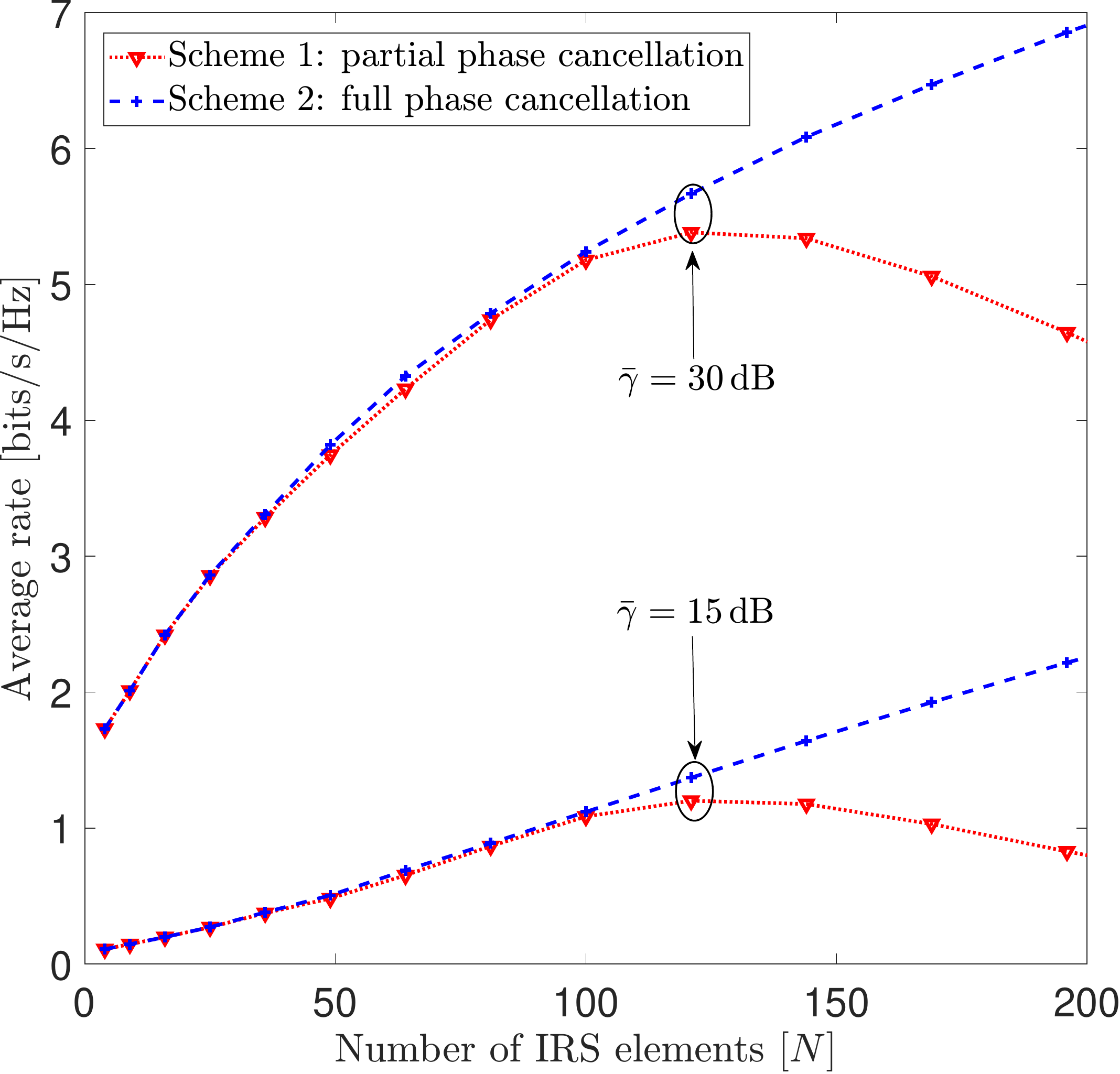}\vspace{-2mm}
	\caption{Impact of spatial correlation over Rician fading.   The Rician factors for $v$, $\mathbf g$, and $\mathbf h$ channels are given by $K_v=2$, $K_g=3$, and $K_h=4$, respectively. The area of IRS is $1\times 1$\,m$^2$, $d_{SD}=100$\,m,    $d_{SI}=d_{DI}=80$\,m, and $\lambda=0.1$\,m. }
	\label{fig:RateCorrelation}\vspace{-10mm}
\end{figure}

In Fig. \ref{fig:RateCorrelation}, the  impact of spatially correlated fading is investigated  by plotting the average achievable rate as a function of the number of reflective elements ($N$) when the surface area of IRS is restricted to $1\times 1$\,m$^2$.  The   co-phasing of received signals at $D$ is achieved by  via (\ref{eqn:scheme2}) by considering the correlation matrices at the IRS, and the corresponding rate curves are denoted by  Scheme-2. To reveal the detrimental impact of spatial correlation, 
 the rate curves without considering the additional phases introduced by correlation are also plotted and the underlying curves are denoted by Scheme-1. 
The achievable rates of the Scheme-2 clearly outperform {Scheme-1} when $N$ grows large. This is because, the impact of correlation {becomes} more severe when a large number of reflective elements are embedded in a given surface area, and the  Scheme-2 is designed to cancel the correlation effects, whereas in Scheme-1, the spatial correlation matrices at the  IRS  are not considered in designing IRS phase-shifts. Thus, in Scheme-2, the received signals via the direct and reflected channels can be perfectly co-phased,  while the same is not possible in Scheme-1. This observation reveals that the deleterious  impact of  spatial correlation can be fully mitigated by intelligently controlling the IRS phase-shifts when the spatial correlation matrices are known.

\vspace{-0mm}
\section{Conclusion}\label{sec:conclusion}

In this paper, the performance of IRS-assisted wireless communication system over Nakagami-$m$ fading  has been investigated. The optimal IRS phase-shifts to maximize the received  SNR have been quantified, and thereby, the probability distributions of this optimal SNR have been  approximated by using the {CLT}. The fundamental performance bounds, including    tight approximations to the outage probability, average SER, and achievable rate bounds have been derived in closed-form.   The accuracy of these performance metrics has been validated via numerical results, and it has been revealed that the tightness of our approximations/bounds improves  when the number of IRS reflective elements   grows large.
The achievable diversity order has been quantified by deriving a  {single}-polynomial high SNR  approximation of the CDF of the optimal  SNR. Thereby, we reveal that the  diversity order can be increased  linearly with  the number of passive IRS reflective elements, and this diversity gain is achieved without deploying additional RF chains in the system. 
It has been shown that our lower/upper rate bounds become asymptotically exact in large $N$ regime, and the transmit power can be scaled inversely proportional to $N^2$.  
The deleterious effects of quantized IRS phase-shifts have been investigated by deriving the   achievable rate bounds, and consequently, it has been revealed that the rate losses due to phase-shift quantization errors can be circumvented in the large IRS reflective element regime.  Further, the deleterious impact of spatially correlated fading has been investigated.
By virtue of  our analysis and numerical results, we show that the performance of wireless communication systems can be significantly improved by coating physical objects in  the propagation channel with the IRS having passive reflective elements and intelligently controlling their phase-shifts.


\appendices

\section{Proof for Lemma {2}} \label{app:AppendixA1} 
\vspace{-3mm}

The RVs $\bar{h}_n$ and $\bar{g}_n$ for $n\in \{1,\cdots, N\}$ in (\ref{eqn:SNR_Optimal}) are independently distributed. Next, we define $\tilde W_n=\eta_n \bar{g}_n \bar{h}_n$ such that it  generates a sequence of independent RVs   $\{\tilde W_1, \cdots, \tilde W_n, \cdots, \tilde W_N\}$; each with mean {$\mu_{\tilde W_n}$ } and variance $\sigma^2_{ \tilde W_n}$. Specifically,  $\mu_{\tilde W_n}$ and $\sigma^2_{ \tilde W_n}$ can be derived as 
\begin{eqnarray}
\mu_{\tilde W_n}&=&\E{\eta_n \bar{g}_n \bar{h}_n}\stackrel{(a)}{=}\eta_n \E{\bar{g}_n} \E{\bar{h}_n}
=\eta_n\sqrt{\frac{\kappa_{g_n}\kappa_{h_n}}{m_gm_h}}T(m_g,m_h,1/2),\\
\sigma^2_{\tilde W_n}&=&\E{(\eta_n \bar{g}_n \bar{h}_n)^2}-\mu^2_{W_n}\stackrel{(b)}{=}\eta^2_n\E{ \bar{g}^2_n} \E{\bar{h}^2_n}-\mu^2_{W_n}\stackrel{(c)}{=}\eta^2_n\kappa_{g_n}\kappa_{h_n}-\mu^2_{\tilde W_n},\label{eqn:variance}
\end{eqnarray}
where $T(a,b,i)$ is defined in Lemma {2}.
In   steps $(a)$ and $(b)$, the fact that $\bar{g}_n$ and  $\bar{h}_n$ are independent RVs is invoked. The square of a Nakagami-$m$ distributed RV follows a Gamma distribution, i.e., $\bar{a}^2_n\sim \mathrm{Gamma}(m_a,\kappa_{a_n}/m_a)$, and hence, in the step $(c)$, we invoke $\E{\bar{a}^2_n}=\kappa_{a_n}$ for $\bar{a}\in\{\bar{g},\bar{h}\}$ \cite{papoulis02}. 
The distribution of $\tilde W=\sum_{n=1}^N \tilde W_n $ approaches to a lower-tail truncated normal distribution if it satisfies the following sufficient conditions \cite{Papoulis2002}; (i) $\sum_{n=1}^N\sigma^2_{\tilde W_n}\stackrel{N \rightarrow \infty}{\rightarrow} \infty,$ and (ii) there exists a number $\alpha>2$ and a finite constant $M$ such that $\E{\tilde W^\alpha_n}<M<\infty$ for all $n\in \{1, \cdots, N\}$. It can be  shown that $\tilde W$ satisfies the first condition  because of the fact that
\vspace{-2mm}
\begin{eqnarray}\label{eqn:condition1}
\lim\limits_{N\rightarrow\infty}\sum_{n=1}^N\sigma^2_{\tilde W_n} &=&\left(1-\frac{T^2(m_g,m_h,1/2)}{m_gm_h}\right) \left[\lim\limits_{N\rightarrow\infty}\sum_{n=1}^N\eta^2_n\kappa_{g_n}\kappa_{h_n}\right]
\stackrel{(d)}{\rightarrow} \infty,
\end{eqnarray} 
\vspace{-7mm}

\noindent
where the step $(d)$ is possible since $\eta^2_n\kappa_{g_n}\kappa_{h_n}>0$, $\forall n$, and as $N$ grows large, $\sum_{n=1}^N\eta^2_n\kappa_{g_n}\kappa_{h_n}$ approaches infinity. 
Next, via the steps similar to  (\ref{eqn:variance}),   $\mathbb E[{\tilde W^\alpha_n}]$ can be computed as 
\begin{eqnarray}\label{eqn:xalpha}
\E{\tilde W^\alpha_n}&=&\eta^\alpha_n\E{ \bar{g}^\alpha_n}\E{ \bar{h}^\alpha_n} {=}  T(m_g,m_h,\alpha/2)\left[\frac{\eta^2_n}{\kappa_{g_n}\kappa_{h_n}}\right]^{\alpha/2}.
\end{eqnarray}
Thus, there exists a finite $M$ such that $\mathbb E[{\tilde W^\alpha_n}]<M<\infty$, $\forall n$ for some $\alpha$, where $T(m_g,m_h,\alpha/2)$ $\times[{\eta^2_n}\big/{\kappa_{g_n}\kappa_{h_n}}]^{\alpha/2}<M$. This completes the proof that $W$ satisfies sufficient conditions, and thus,  $F_W(w)$ can be approximated by a lower-tail truncated  normal  distribution   as stated in Lemma {2} with $\mu_W$ and $\sigma^2_W$, which are given in (\ref{eqn:meanS}).

\vspace{-2mm}

\section{Derivation of the CDF of $\gamma$ in (\ref{eqn:cdf_SNR})} \label{app:AppendixBB} 
\vspace{-2mm}

To begin with, by defining the received signal envelope as $R=\bar{v}+ W$, we   rewrite (\ref{eqn:SNR_Optimal}) as $\gamma=\bar{\gamma}R^2$.  
Since, $\bar{v}$ and $W$ are independent RVs, the PDF of $R$ can be derived as \cite{Papoulis2002}  

\vspace{-10mm}
\begin{eqnarray}\label{eqn:Apx_1_eqn_1}
f_R(r) &=& \int_{0}^{\infty} f_{\bar{v}}(u) f_W(r-u) du 
= 2a^{m_v} \lambda \exp{-\frac{(r-\bar{\mu})^2}{2\bar{\sigma}^2}} \int_{0}^{\infty} u^{\tilde m_v} \exp{-au^2+ub(r)} du \nonumber \\
&=& 2a^{m_v} \lambda \exp{-\frac{(r-\bar{\mu})^2}{2\bar{\sigma}^2}} \exp{\frac{b^2(r)}{4a}}  \underbrace{\int_{0}^{\infty} u^{\tilde m_v} \exp{-a \left(u-\frac{b(r)}{2{a}}\right)^2} du}_{\mathrm I_R(r)},
\end{eqnarray}  
\vspace{-6mm}

\noindent where $\tilde m_v=2m_v-1$, and $b(r)$ denotes a function of $r$ defined  by $b(r) = (r-\bar{\mu})/\bar{\sigma}^2$. In (\ref{eqn:Apx_1_eqn_1}), $a$ and $\lambda$ are defined in \eqref{eqn:def_a}. Substituting $t=\sqrt{a} (u-{b(r)}/{2{a}})$, $\mathrm I_R(r)$ in  \eqref{eqn:Apx_1_eqn_1} can be rewritten as
\vspace{-2mm}
\begin{eqnarray}\label{eqn:Apx_1_eqn_2}
\mathrm I_R(r)&=& \frac{1}{a^{m_v}} \int_{-\tilde b(r)}^{\infty} \left(t+{\tilde b(r)}\right)^{\tilde m_v} \exp{-t^2} dt = \frac{1}{a^{m_v}} \sum_{k=0}^{\tilde m_v} \binom{\tilde m_v}{k} \left({\tilde b(r)}\right)^{\tilde m_v-k} \mathcal I(k,-\tilde b(r)),  
\end{eqnarray}  
\vspace{-8mm}

\noindent where $\tilde b(r)=(r-\bar{\mu})/(2\bar{\sigma}^2\sqrt{a})$ and $\mathcal I(q,x)=\int_{x}^{\infty} t^{q} \exp{-t^2} dt$. For $x\geq 0$, $\mathcal I(q,x)=\Gamma((q+1)/2,x^2)/2$, and when  $x<0$, $\mathcal I(q,x)$ can be written as $\mathcal I(q,x)=\int_{-\infty}^{\infty} t^{q} \exp{-t^2} dt-\int_{-\infty}^{-x} t^{q} \exp{-t^2} dt$, which can be evaluated in closed-form as   (\ref{eqn:Ikx}).
By substituting $\mathcal I(k,-\tilde b(r))$ into (\ref{eqn:Apx_1_eqn_2}),  $\mathrm I_R(r)$ can be evaluated, and then, by using the resulting expression, $f_R(r)$ in (\ref{eqn:Apx_1_eqn_1}) can be derived as 
\vspace{-1mm}
\begin{eqnarray}\label{eqn:pdf_R}
\!\!\!\!\!\!\!\!\!\!\!f_R(r) \!&=&\! 
\begin{cases}\!
\lambda \exp{-\Delta \left(\frac{r-\bar{\mu}}{2 \sqrt{a}\bar{\sigma}^2}\right)^2} \!\!\sum_{k=0}^{\tilde m_v}   \binom{\tilde m_v}{k}  \left(\frac{r-\!\bar{\mu}}{2\bar{\sigma}^2 \sqrt{a}}\right)^{\!\tilde m_v-\!k}
\Gamma \!\left(\!\frac{k\!+\!1}{2}, \!\left(\!\frac{(r\!-\!\bar{\mu})}{2\bar{\sigma}^2 \sqrt{a}}\right)^2\right) &r\leq \bar{\mu}, \\\!
\lambda \exp{-\Delta \left(\frac{r-\bar{\mu}}{2 \sqrt{a}\bar{\sigma}^2}\right)^2} \!\!\sum_{k=0}^{\tilde m_v}   \binom{\tilde m_v}{k}  \left(\frac{r-\!\bar{\mu}}{2\bar{\sigma}^2 \sqrt{a}}\right)^{\!\tilde m_v-\!k}
\!\!\left[\Gamma\left(\frac{k+1}{2}\right)\!+\!(-1)^k\gamma \!\left(\!\frac{k\!+\!1}{2}, \!\left(\!\frac{(r\!-\!\bar{\mu})}{2\bar{\sigma}^2 \sqrt{a}}\right)^2\right)\right] &r> \bar{\mu},
\end{cases}
\!\!
\end{eqnarray}
and $f_R(r)=0$ otherwise.

Next, the derivation of the  CDF of $R$ is outlined. For $0<r\leq \bar{\mu}$, the CDF of $R$ can be written as $F_R(r)=\int_0^r f_R(y) dy$. By substituting \eqref{eqn:pdf_R},  $F_R(r)$ can be written as 
\begin{eqnarray}\label{eqn:cdf_R1}
F_{R}(r) =\sum_{k=0}^{\tilde m_v}2\lambda\sqrt{a}\bar{\sigma}^2 \binom{\tilde m_v}{k} \mathrm J_1, \text{ for } 0<r\leq \bar{\mu}.\;\;
\end{eqnarray}
In (\ref{eqn:cdf_R1}), $\mathrm J_1$ is given by
\begin{eqnarray}\label{eqn:J}
\mathrm J_1\!=\!\int_0^r \!\!\frac{(\tilde b(y))^{\tilde m_v-k}\exp{-\Delta (\tilde b(y))^2}}{2\sqrt{a}\bar{\sigma}^2}\Gamma\left(\!\frac{k+1}{2},(\tilde b(y))^2\!\right)dy.\quad 
\end{eqnarray}
By substituting $t=-\tilde b(y)$, $\mathrm J_1$ can be rewritten as 
\begin{eqnarray}\label{eqn:J2}
\mathrm J_1=(-1)^{\tilde m_v-k}\left[\mathcal J(k,-\tilde b(0))-\mathcal J(k,-\tilde b(r))\right],
\end{eqnarray}
where $\mathcal J(k,\delta)$ is given by 
\begin{eqnarray}\label{eqn:mathcalJ}
\mathcal J\left(k,z\right)=\int_z^\infty t^{\tilde m_v-k}\exp{-\Delta t^2}\Gamma\left(\frac{k+1}{2},t^2\right)dt.
\end{eqnarray}
Next,   (\ref{eqn:mathcalJ}) can be evaluated separately for odd and even values of $k$ as follows: When $k$ is odd,  $\delta_o={(k+1)}/{2}$ becomes an integer, and we  expand $\Gamma({(k+1)}/{2},t^2)$   in (\ref{eqn:mathcalJ}) as $\Gamma({(k+1)}/{2},t^2)=(\delta_o-1)! \exp{-t^2} \sum\nolimits_{i=0}^{\delta_o-1} t^{2i}/i!$ \cite[8.352.7]{Gradshteyn2007}. For odd values of $k$, $\mathcal J(k,z)$ can be  rewritten as
\begin{eqnarray}\label{eqn:Jodd}
\mathcal J_o\left(k,z\right)=(\delta_o-1)!\sum_{i=0}^{\delta_o-1}\frac{1}{i!} \int_z^\infty   t^{\tilde m_v-k+2i}\exp{-(\Delta+1) t^2} dt.\quad
\end{eqnarray}
By using \cite[2.33.10]{Gradshteyn2007}, $\mathcal J_o\left(k,z\right)$ can be evaluated in closed-form as   (\ref{eqn:I_mu_odd}). For even values of $k$, $\delta_e=m_v-k/2$ becomes an integer, and we have {\cite[2.33.11]{Gradshteyn2007} }
\begin{eqnarray}\label{eqn:evencase}
\int t^{\tilde m_v-k}\exp{-\Delta t^2}dt = \frac{(\delta_e-1)!}{2} \sum_{j=0}^{\delta_e-1} -\frac{\exp{-\Delta t^2} t^{2j}}{j! \Delta^{\delta_e-1}}.
\end{eqnarray}
Thus, $\mathcal J\left(k,z\right)$ can be evaluated by using integration-by-parts as follows:  
\begin{eqnarray}\label{eqn:Apx_2_eqn_6}
\!\!\!\!\!\!\!\!\!\!\!\mathcal J_e(k,z) 
&=&\frac{(\delta_e-1)!}{2} \sum_{j=0}^{\delta_e-1}\frac{\exp{-\Delta z^2} z^{2j}}{j! \Delta^{\delta_e-1}} \Gamma\left(\frac{k+1}{2}, z^2\right)
\!-\! (\delta_e-1)! \sum_{j=0}^{\delta_e-1} \frac{\Delta^{\delta_e-1}}{j!} \!\!\int_{z}^{\infty} \!\!\!t^{k+2j} \exp{-(\Delta+1)t^2}dt.
\end{eqnarray} 
By invoking \cite[2.33.10]{Gradshteyn2007}, $\mathcal J_e\left(k,z\right)$ can be evaluated in closed-form as   (\ref{eqn:I_mu_even1}). Depending on the value of $k$, by substituting (\ref{eqn:I_mu_odd}) or (\ref{eqn:I_mu_even1}) into (\ref{eqn:J2}), the desired expression for $\mathrm J_1$ can be derived  next. For $r> \bar{\mu}$,  $F_R(r)=1-\int_r^\infty f_R(y) dy$. By substituting \eqref{eqn:pdf_R},  $F_R(r)$ can be written as 
\begin{eqnarray}\label{eqn:cdf_R2}
F_{R}(r) &=&1-\sum_{k=0}^{\tilde m_v}2\lambda\sqrt{a}\bar{\sigma}^2\binom{\tilde m_v}{k}  \left(\mathrm I_2-\mathrm J_2\right) \text{ for } r> \bar{\mu},\qquad
\end{eqnarray}
where $\mathrm I_2$ and $\mathrm J_2$ can be defined as 
\begin{eqnarray}
\mathrm I_2\!&=&\!\frac{[(-1)^k+1]\Gamma\left(\frac{k+1}{2}\right)}{2\sqrt{a}\bar{\sigma}^2}\int_r^\infty \left(\tilde b(y)\right)^{\tilde m_v-k}\exp{-\Delta(\tilde b(y))^2}dy,\quad\;\;\\
\mathrm J_2\!&=&\!(-1)^k\!\!\int_r^\infty \!\frac{\left(\tilde b(y)\right)^{\tilde m_v-k}\exp{-\Delta(\tilde b(y))^2}}{2\sqrt{a}\bar{\sigma}^2}\Gamma\!\left(\!\frac{k+1}{2},(\tilde b(y))^2\right)dy.\quad\;\;
\end{eqnarray}
By substituting $t=\tilde b(y)$ and invoking \cite[2.33.10]{Gradshteyn2007}, $\mathrm I_2$ can be derived as 
\begin{eqnarray}\label{eqn:I2}
\mathrm I_2=\frac{[(-1)^k+1]\Gamma\left(\frac{k+1}{2}\right)}{2}\Gamma\left(\frac{\tilde m_v-k+1}{2},\Delta(\tilde b(r))^2 \right).\quad
\end{eqnarray}
By following    manipulations similar to those in (\ref{eqn:Jodd}) and(\ref{eqn:Apx_2_eqn_6}), $\mathrm J_2$ can be derived as 
\begin{eqnarray}\label{eqn:J_2}
\mathrm J_2={(-1)^k\mathcal J(k,\tilde b(r))},
\end{eqnarray}
where $\mathcal J(k,\tilde b(r))$ is defined in (\ref{eqn:Jkx}). By substituting (\ref{eqn:I2}) and (\ref{eqn:J_2}) into (\ref{eqn:cdf_R2}), $F_{R}(r)$ can be derived for $r> \bar{\mu}$. From probability theory, for ${\gamma}=\bar{\gamma}R  ^2$ with $\bar{\gamma}>0, R >0$, we can write the PDF and CDF of $\gamma$ as $f_{\gamma}(y)=f_{R}(\sqrt{y/\bar{\gamma}})/(2\sqrt{y\bar{\gamma}})$ and  $F_{\gamma}(y)=F_{R}(\sqrt{y/\bar{\gamma}})$, respectively \cite{papoulis02}. By using this fact, $F_{\gamma}(\cdot)$ can be written as   \eqref{eqn:cdf_SNR}.

\vspace{-2mm}

\section{Derivation of diversity order in (\ref{eqn:diversity}) {and asymptotic SER in (\ref{eqn:asympSER})}}\label{app:AppendixCC} 
We notice that $\tilde W_n=\eta_n \bar{g}_n\bar{h}_n$ is a scaled product of two Nakagami-m RVs. Hence, the PDF of {$\tilde W_n$} is given by \cite{Karagiannidis2007} 
\vspace{-3mm}
\begin{eqnarray}\label{eqn:PDFXm}
f_{{\tilde W_n}}(\omega)=\psi_n\omega^{m_a+m_b-1}K_{m_a-m_b}\left(\omega \tau_n\right) \text{ for } \omega \geq 0,
\end{eqnarray}
where 
$\psi_n\!=\!4(m_am_b)^{(m_a+m_b)/2}(\eta_n\sqrt{\kappa_{a}\kappa_b})^{-(m_a+m_b)/2}\!/(\Gamma(m_a)\Gamma(m_b))$ and $\tau_n\!=\!2\sqrt{m_am_b/\!(\kappa_a\kappa_b \eta^2_n)}$.  In (\ref{eqn:PDFXm}),  $K_v(\cdot)$ is the modified Bessel function of the second kind  \cite[6.624.1]{Gradshteyn2007}. The moment generating function (MGF) of  $\tilde W_n$ can be derived via \cite[6.621.3]{Gradshteyn2007} as follows:
\begin{eqnarray}
\mathcal M_{\tilde W_n}(s)&=& \psi'_n (s+\tau_n)^{-2m_a}\mathcal F\left(2m_a, m_a-m_b+{1}/{2};m_a+m_b+{1}/{2};{s-\tau_n}/({s+\tau_n})\right),
\end{eqnarray}
where $ \psi'_n=\sqrt{\pi}\psi_n(2\tau_n)^{m_a-m_b}\Gamma(2m_a)\Gamma(2m_b)/\Gamma(m_a+m_b+1/2)$ and $\mathcal F(\cdot,\cdot;\cdot;\cdot)$ is the hyper-geometric function \cite[9.100]{Gradshteyn2007}. Next, by utilizing the fact that  $\{\tilde W_1,\tilde W_2\cdots,\tilde W_n\cdots, \tilde W_N\}$ are independent RVs, the MGF of $\tilde W=\sum_{n=1}^N \tilde W_n$ can be derived as $\mathcal M_{\tilde W}(s)=\prod_{n=1}^{N}\mathcal M_{\tilde W_n}(s)$ \cite{Papoulis2002}. The order of smoothness of the PDF of {$\tilde W$} at the origin can be quantified by using the  decaying order of the MGF $\mathcal M_{\tilde W}(s)$ \cite{Wang2003a}. To this end, via the fact that ${(s-\tau_n)}/{(s+\tau_n)}\rightarrow 1$ as $s\rightarrow \infty$, and since $m_a<m_b$ satisfies the condition \cite[9.122.1]{Gradshteyn2007} when $s\rightarrow \infty$, $\mathcal M_{ \tilde W}(s)$ can be asymptotically approximated as
\begin{eqnarray}\label{eqn:asymptoticMGF}
\mathcal M^{\infty}_{\tilde W}(s)=\left[\prod_{n=1}^N\tilde{\psi}_n\right] s^{-2m_a N},
\end{eqnarray}
where  $\tilde{\psi}_n={\psi}_n\Gamma(m_a+m_b+1/2)\Gamma(2m_b-2m_1-1)/(\Gamma(m_b-m_a+1/2)\Gamma(2m_b))$. 

Next, the behavior of  the PDF of the SNR pertaining to the direct channel in high SNR regime is governed by the behavior of $f_{\bar{V}}(x)$ around $x=0$. By substituting the Maclaurin series expansions of the exponential functions, the PDF can be approximated near the origin as $f^{ 0^+}_{\bar{V}}(x)=\psi'_vx^{2m_v-1}+\mathcal O(x^{2m_v})$, where $\psi'_v=2m_v^{m_v}/(\Gamma(m_v)\kappa^{m_v}_v)$. The corresponding MGF can be derived as \cite{papoulis02}
\begin{eqnarray}\label{eqn:asymptoticMGFdirect}
\mathcal M^{\infty}_{\bar{V}}(s)=\psi_v/(s^{2m_v}) \text{ and } \psi_v=\Gamma(2m_v)\psi'_v.
\end{eqnarray}
Since the direct and  reflected channels via the IRS are independent RVs, the MGF of $r=v+{\tilde W}$ when $s\rightarrow \infty$ can be derived by using (\ref{eqn:asymptoticMGF}) and (\ref{eqn:asymptoticMGFdirect}) as follows:
\begin{eqnarray}\label{eqn:asymptoticMGFR}
\mathcal M^{\infty}_{R}(s)&=&\mathcal M^{\infty}_{\bar{V}}(s)\mathcal M^{\infty}_{ \tilde W}(s)=\psi_v\left[\prod_{n=1}^N\tilde{\psi}_n\right]\Big/(s^{2m_v+2m_aN}).
\end{eqnarray}
By taking inverse Laplace transform of (\ref{eqn:asymptoticMGFR}), the PDF of $R$ can be approximated by a single polynomial term for $r\rightarrow 0^+$   as 
\begin{eqnarray}\label{eqn:pdfR}
f^{ 0^+}_{R}(r)=\Omega_R {r^{2m_v+2m_a N-1}} +\mathcal O(r^{2m_v+2m_a N}), 
\end{eqnarray}
where ${\Omega_R}=\psi_v\left[\prod_{n=1}^N\tilde{\psi}_n\right]/\Gamma(2m_v+2m_aN)$. From (\ref{eqn:pdfR}),  as $\bar \gamma \rightarrow \infty$, the    CDF can be approximated at the origin as  $F^{0^+}_R(r)=\int_{0}^{r}f^{0^+}_R(t)dt$. By performing  the variable transformation $r=\sqrt{y/\bar{\gamma}}$ \cite{Papoulis2002}, a single-polynomial asymptotic  approximation of the CDF of $\gamma$ is derived as
\begin{eqnarray}\label{eqn:cdfgamma}
F^{ 0^+}_{\gamma}({y})=\Omega_{op} {\left({y}\big/{\bar{\gamma}}\right)^{G_d}} +\mathcal O\left(\left({y}\big/{\bar{\gamma}}\right)^{G_d+1}\right), 
\end{eqnarray}
where $G_d$ and $\Omega_{op}$ are defined in (\ref{eqn:diversity}) and (\ref{eqn:omegaop}), respectively. Then, the asymptotic outage probability and the diversity order can be derived as  (\ref{eqn:Poutasym}) and  (\ref{eqn:diversity}), respectively.

Next, the derivation of asymptotic average  SER in the high SNR regime  (\ref{eqn:asympSER})  is outlined. To begin with, the  average SER can be written  in an integral form as \cite{Amarasuriya2010b}
\begin{eqnarray}\label{eqn:PE}
 \bar{P}_e=\E{\alpha\Q{\sqrt{\beta\tilde \gamma}}} = \alpha\sqrt{\beta}/(2\sqrt{2\pi})\int^{\infty}_{0}y^{-1/2}\mathrm{exp}(-\beta y/2)F_{\gamma}(y)dy
\end{eqnarray}
By substituting (\ref{eqn:cdfgamma}) into $\bar{P}_e$ (\ref{eqn:PE}), an asymptotic    average SER can be written  as
	\begin{eqnarray}\label{eqn:asympPe}
	\bar{P}^{\infty}_e=\frac{\alpha\sqrt{\beta}\Omega_{op}}{2\sqrt{2\pi}\bar{\gamma}^{G_d}}\int_{0}^\infty y^{G_d-\frac{1}{2}}\mathrm{exp}(-\beta y/2) dy.
	\end{eqnarray}
By substituting $t=\beta y/2$ into (\ref{eqn:asympPe}), and evaluating the integral via \cite[Eqn. (8.310.1)]{Gradshteyn2007}, the asymptotic  average SER in  high SNR regime can be derived as (\ref{eqn:asympSER}).

\vspace{-4mm}

\section{Derivation of $\mathcal R_{lb}$ in (\ref{eqn:rate_lb_sub}), $\mathcal R_{ub}$ in (\ref{eqn:rate_ub_sub}), and $\mathcal R^{\infty}$ in (\ref{eqn:asymptoticrate})} \label{app:AppendixDD} 

To begin with, the expectation term in \eqref{eqn:rate_ub} can be simplified as 
\begin{eqnarray}\label{eqn:E_gamma_ub}
\E{\gamma} = \bar{\gamma} \left(\E{\bar{v}^2}+2\E{\bar{v}}\E{W}+ \E{W^2} \right).
\end{eqnarray} 
The RVs in (\ref{eqn:E_gamma_ub}) are distributed  as $\bar{v}^2\sim \mathrm{Gamma}(m_v,\kappa_{v}/m_v)$, $\bar{v}\sim \mathrm{Nakagami}(m_v,\kappa_{v})$ and $W\sim \mathcal{N}^+(\mu_W,\sigma^2_W)$. Thus, the expectation terms in (\ref{eqn:E_gamma_ub}) can be derived as \cite{papoulis02}
\begin{eqnarray}
\!\!\!\!\!\!\!\E{\bar{v}^2}= \kappa_v, \;\; \E{\bar{v}}=\frac{\Gamma\left(m_v + 1/2\right)}{\Gamma\left(m_v\right)}  \left({\frac{\kappa_v}{m_v}}\right)^{1/2},\label{eqn:meanv}
\E{W}=\mu_W, \text{ and }\E{W^2} = \mu^2_W+\sigma^2_W.\label{eqn:meanS2}
\end{eqnarray}
By substituting (\ref{eqn:meanv}) into (\ref{eqn:E_gamma_ub}), and replacing $\E{\gamma} $ in (\ref{eqn:rate_ub}) with (\ref{eqn:E_gamma_ub}),  $\mathcal{R}_{ub}$ can be derived in closed-form as   (\ref{eqn:rate_ub_sub}).

Next, we outline the  derivation of a lower bound for the average achievable rate. To begin with, by applying the Taylor series expansion for $1/\gamma$ around $\E{\gamma}$,   the term $\E{1/\gamma}$ in (\ref{eqn:rate_lb})  can be approximated as \cite{Zhang2014}
\begin{eqnarray}\label{eqn:E_gamma_lb}
\E{1/\gamma}\approx {1}\big/{\E{\gamma}} + {\Var{\gamma}}\big/{\left(\E{\gamma}\right)^3}={\E{\gamma^2}}\big/{\left(\E{\gamma}\right)^3},
\end{eqnarray}
where $\E{\gamma}$ is defined in \eqref{eqn:E_gamma_ub} and  $\E{\gamma^2}$ is evaluated using $\E{\gamma^2} = \bar{\gamma}^2\E{(\bar{v}+W)^4}$. Since $\bar{v}$ and $W$ are independent RVs, we have $\E{(\bar{v}+W)^4} = \E{\bar{v}^4} + 4\E{\bar{v}^3}\E{W} + 6\E{\bar{v}^2}\E{W^2} + 4\E{\bar{v}}\E{W^3} +\E{W^4}$. By using the fact that $\bar{v}^2\sim \mathrm {Gamma}(m_v,\kappa_{v}/m_v)$ and invoking steps similar to those used in deriving (\ref{eqn:xalpha}), $\E{\bar{v}^\alpha}$ for $\alpha\in\{1,2,3,4\}$ can be derived  as $\E{\bar{v}^\alpha}={\Gamma\left(m_v+\alpha/2\right)}\left({\kappa_v}/{m_v}\right)^{\alpha/2}\big/{\Gamma\left(m_v\right)}$.
Furthermore, by following mathematical manipulations that are used in deriving (\ref{eqn:pdf_R}), $\E{W^\alpha}$ for $\alpha \in \{1,2,3,4\}$ can be written as
\begin{eqnarray}\label{eqn:E_S_n}
\E{W^\alpha} = \frac{\xi}{2\sqrt{\pi}} \sum_{i=0}^{\alpha} \! \binom{\alpha}{i} \!\left( \!\sqrt{2 \bar{\sigma}^2}\right)^{i} \!\! \bar{\mu}^{\alpha-i} \mathcal I\!\left(i, \frac{-\bar{\mu}}{\sqrt{2 \bar{\sigma}^2}}\right), 
\end{eqnarray}
where  $\mathcal I(\cdot,\cdot)$ is given in (\ref{eqn:Ikx}).
By using $\E{\bar{v}^\alpha}$ and $\E{W^\alpha}$ for $\alpha \in \{1,2,3,4\}$,  $\E{(\bar{v}+W)^4}$ can be computed. Then, by substituting $\E{\gamma}$ and $\E{\gamma^2}$ into (\ref{eqn:E_gamma_lb}), $\E{1/\gamma}$ can be computed. Thereby, by substituting $\E{1/\gamma}$ into (\ref{eqn:rate_lb}), $\mathcal{R}_{lb}$ can be derived as in \eqref{eqn:rate_lb_sub}.

{
Next, the derivation of the asymptotic  rate in (\ref{eqn:asymptoticrate}) is outlined. By assuming $\kappa_{g_n}=\kappa_{g}$, $\kappa_{h_n}=\kappa_{h}$ and $\eta_{n}=\eta$ $\forall n$, the variables $\bar{\mu}$ and $\bar{\sigma}^2$ can be simplified as $\bar{\mu}=N\bar{\mu}_{\infty}$ and $\bar{\sigma}^2=N\bar{\sigma}^2_{\infty}$, respectively, where $\bar{\mu}_{\infty}$ and $\bar{\sigma}^2_{\infty}$ can be written as $\bar{\mu}_{\infty}=\eta\sqrt{{\kappa_{g}\kappa_{h}}\big/{m_gm_h}}T(m_g,m_h,1/2),\label{eqn:meanbarinf}$
and $\bar{\sigma}_{\infty}^2=\eta^2\kappa_{g}\kappa_{h}\left(1-{T^2(m_g,m_h,1/2)}\big/{m_gm_h}\right)$,
When $N\rightarrow \infty$,  the SNR    (\ref{eqn:rate_lb_sub}) can be derived as 
\begin{eqnarray}\label{eqn:Apx_3_3_eqn_3}
\!\!\!\!\!\!\lim_{N \rightarrow \infty} \gamma_{lb} &=& \lim_{N \rightarrow \infty} \frac{\bar{\gamma}N^2 \left(\frac{\kappa_v}{N^2} + 2 \frac{\mu_{W}}{N^2}{\Gamma\left(m_v + 1/2\right)}\sqrt{{\kappa_v}/{m_v}}/{\Gamma\left(m_v\right)}  +\frac{\mu^2_W}{N^2} +\frac{\sigma_{W}^2}{N^2} \right)^3}
{\left(\frac{\xi}{2\sqrt{\pi}} \bar{\mu}_{\infty}^{4} \mathcal I\left(0, \frac{-\sqrt{N}\bar{\mu}_{\infty}}{\sqrt{2 \bar{\sigma}^2_{\infty}}}\right)+\frac{\Lambda(N)}{N^4}\right) }
\stackrel{(a)}{=}\bar{\gamma}_E\bar{\mu}_{\infty}^2,\;\;\;
\end{eqnarray}
where $\bar{\gamma}_E=\lim_{N \rightarrow \infty}\bar{\gamma}/N^2$, $\mathcal I (\cdot,\cdot)$ is defined in (\ref{eqn:Ikx}), $\Lambda(N)$ is a polynomial of order $3$ and it captures the remaining terms of the expansion of the denominator of $\gamma_{lb}$. The limits in step $(a)$ of (\ref{eqn:Apx_3_3_eqn_3}) are evaluated by invoking (\ref{eqn:asympresult}), $ \lim_{N \rightarrow \infty}\xi= 1$, $\lim_{N \rightarrow \infty}\mathcal I\left(0, {-\sqrt{N}\bar{\mu}_{\infty}}\big/{\sqrt{2 \bar{\sigma}^2_{\infty}}}\right)=\mathcal I\left(0, -\infty\right)=2\gamma(1/2)$ and $\lim_{N \rightarrow \infty} {\mu^2_W}\big/{N^2}=\bar{\mu}_{\infty}^2$. Similarly, in the asymptotic regime, the SNR term in $\mathcal R_{ub}$ (\ref{eqn:rate_ub_sub}) can be derived as 
\begin{eqnarray}\label{eqn:Apx_3_3_eqn_4}
\lim_{N \rightarrow \infty} \gamma_{ub} &=& \lim_{N \rightarrow \infty} {\bar{\gamma}N^2 \left(\frac{\kappa_v}{N^2} + 2 \frac{\mu_{W}}{N^2}{\Gamma\left(m_v + 1/2\right)}\sqrt{{\kappa_v}/{m_v}}/{\Gamma\left(m_v\right)}  +\frac{\mu^2_W}{N^2} +\frac{\sigma_{W}^2}{N^2} \right)}
{=}\bar{\gamma}_E\bar{\mu}_{\infty}^2.\;\;\;\;\;\;
\end{eqnarray}
Thus, we have $\lim_{N \rightarrow \infty}  \mathcal R_{lb}=\lim_{N \rightarrow \infty}  \mathcal R_{ub}=\log{1+\bar{\gamma}_E\bar{\mu}_{\infty}^2}$, which can be rewritten as   (\ref{eqn:asymptoticrate}).
} 

\vspace{-2mm}

\section{The derivation of the average SER in \eqref{eqn:avg_ber}  }\label{app:AppendixEE}

To begin with, $\bar{P_e} $ in (\ref{eqn:PeXX}) can  be alternatively written as  
\begin{eqnarray}\label{eqn:Apx_4_eqn_1}
\bar{P_e} &\approx& \alpha \int_{0}^\infty \int_{0}^\infty \Q{\sqrt{\beta \bar{\gamma}}(x+w)}f_{\bar{v}}(x)f_{W}(w) dxdw ,
\end{eqnarray}
where $f_{\bar{v}}(x)$ and $f_{W}(w)$ are given in (\ref{eqn:Nakagami_v}) and (\ref{eqn:pdf_S}), respectively. The $\mathcal Q$-function,  $\Q{\sqrt{\beta \bar{\gamma}}(x+w)}$, in (\ref{eqn:Apx_4_eqn_1}) can be alternatively written as \cite{Gradshteyn2007}
\begin{eqnarray}\label{eqn:alter_Q}
\Q{\sqrt{\beta \bar{\gamma}}(x+w)}=\frac{1}{\pi}\int_{0}^{\pi/2}\mathrm{exp}\left(-\frac{\beta \bar{\gamma}x^2}{2\sin^2\vartheta}-\frac{\beta \bar{\gamma}y^2}{2\cos^2\vartheta}\right)d\vartheta.
\end{eqnarray}
By substituting (\ref{eqn:alter_Q}) into (\ref{eqn:Apx_4_eqn_1})  and by applying several mathematical manipulations, we have 
\begin{eqnarray}\label{eqn:Pe2}
\bar{P_e} &\approx& \tilde{\mathcal U} \int_{0}^{\pi/2} \mathrm I_{\bar{v}}(\vartheta)  \mathrm I_{W}(\vartheta) d\vartheta,
\end{eqnarray}
where ${\tilde {\mathcal U}}=2m_v^{m_v}\xi \exp{-\bar{\mu}^2/2\bar{\sigma}^2}/\left(\sqrt{2\pi \bar{\sigma}^2}\Gamma(m_v)\kappa_v^{m_v}\pi\right)$. In (\ref{eqn:Pe2}), $\mathrm I_{\bar{v}}(\vartheta)$ and $  \mathrm I_{W}(\vartheta) $  are defined as 
\begin{eqnarray}
\mathrm  I_{\bar{v}}(\vartheta)=\int_0^\infty x^{2m_v-1}\exp{-u_1 x^2}dx \qquad  \text{ and }\qquad 
\mathrm I_{W}(\vartheta)&=&\int_0^\infty \exp{-(z_1 y^2+2z_2y)}dy,
\end{eqnarray}
where $u_1=(m_v/\kappa_v+\beta\bar{\gamma}/(2\sin^2(\vartheta)))$, $z_1=(1/2\bar{\sigma}^2+\beta\bar{\gamma}/(2\cos^2(\vartheta)))$  and $z_2=\bar{\mu}/2\bar{\sigma}^2$. By substituting $t=\sqrt{u_1} x$ and using the definition of Gamma function, $\mathrm I_{\bar{v}}(\vartheta)$ can be evaluated as 
\begin{eqnarray}\label{eqn:Iv}
\mathrm  I_{\bar{v}}(\vartheta)=\frac{1}{2u_1^{m_v}}\int_0^\infty t^{m_v-1}\exp{-t}dt=\frac{\Gamma(m_v)}{2}\left[\frac{m_v}{\kappa_v}+\frac{\beta\bar{\gamma}}{(2\sin^2(\vartheta))}\right]^{-m_v}.
\end{eqnarray}
By invoking \cite[2.33.1]{Gradshteyn2007}, $\mathrm I_{W}(\vartheta)$ can be derived as 
\begin{eqnarray}\label{eqn:Iw}
\mathrm I_{W}(\vartheta)&=&2\sqrt{\frac{\pi}{\left(\frac{1}{2\bar{\sigma}^2}+\frac{\beta\bar{\gamma}}{2\cos^2(\vartheta)}\right)}}
\e{\frac{\bar{\mu}^2/4\bar{\sigma}^4}{\frac{1}{2\bar{\sigma}^2}+\frac{\beta\bar{\gamma}}{2\cos^2(\vartheta)}}}\Q{-\frac{\sqrt{2}\bar{\mu}}{1+\frac{\beta\bar{\gamma}\bar{\sigma}^2}{\cos^2(  \vartheta)}}}.
\end{eqnarray}
By substituting (\ref{eqn:Iv}) and (\ref{eqn:Iw}) into (\ref{eqn:Pe2}), $\bar{P_e}$ can be alternatively written as
\begin{eqnarray}\label{eqn:Pe3}
\bar{P_e} &\approx& \mathcal U \int_{0}^{\pi/2} 
\frac{\e{\bar{\mu}^2\big/\left({{2\bar{\sigma}^2}+\frac{2\beta\bar{\gamma}\bar{\sigma}^4}{\cos^2(\vartheta)}}\right)}}{\left[\frac{m_v}
	{\kappa_v}+\frac{\beta\bar{\gamma}}{(2\sin^2(\vartheta))}\right]^{m_v}\sqrt{\left(\frac{1}{2\bar{\sigma}^2}+\frac{\beta\bar{\gamma}}{2\cos^2(\vartheta)}\right)}}
\Q{-\frac{\sqrt{2}\bar{\mu}}{1+\frac{\beta\bar{\gamma}\bar{\sigma}^2}{\cos^2(\vartheta)}}} d\vartheta,
\end{eqnarray}
where $\mathcal U=\Gamma(m_v)\sqrt{\pi}\tilde{\mathcal U}$. Let $\mathcal P(\vartheta)$ be the integrand of (\ref{eqn:Pe3}), and then, $\vartheta_u$ in (\ref{eqn:vartheta}) provides the argument that maximizes $\mathcal P(\vartheta)$. Therefore, the integral in (\ref{eqn:Pe3}) can be upper bounded as 
\begin{eqnarray}\label{eqn:upper}
 \int_{0}^{\pi/2} \mathcal P(\vartheta)d\vartheta \leq \pi\mathcal P(\vartheta_u)/2.
\end{eqnarray}
By replacing the   integral in  (\ref{eqn:Pe3}) by  (\ref{eqn:upper}), the desired upper bound can be derived as in \eqref{eqn:avg_ber}.


\vspace{-3mm}

  \linespread{1.6}
\bibliographystyle{IEEEtran}
\bibliography{IEEEabrv,References}

\end{document}